\documentclass[12pt,preprint]{aastex}

\shorttitle{M82 H92$\alpha$ Radio Recombination Lines}
\shortauthors{Rodriguez-Rico et al. 2003}

\begin{document}

\title{VLA H92$\alpha$ and H53$\alpha$ Radio Recombination Line Observations of M82}

\author{C. A. Rodriguez-Rico \altaffilmark{1,2}}
\email{crodrigu@nrao.edu}

\author{F. Viallefond\altaffilmark{3}}
\email{fviallef@maat.obspm.fr}

\author{J.-H. Zhao\altaffilmark{4}}
\email{jzhao@cfa.harvard.edu}

\author{W. M. Goss\altaffilmark{1}}
\email{mgoss@nrao.edu}

\author{K. R. Anantharamaiah\altaffilmark{5}}

\altaffiltext{1}{National Radio Astronomy Observatory, Socorro, NM 87801}

\altaffiltext{2}{Centro de Radioastronom\'\i a y Astrof\'\i sica, UNAM, Apdo. Postal 3-72, Morelia, Michoac\'an 58089, M\'exico.}

\altaffiltext{3}{LERMA, Observatoire de Paris, 61 Av. de l'Observatoire F-75014 Paris}

\altaffiltext{4}{Harvard-Smithsonian Center for Astrophysics, 60 Garden Street, Cambridge, MA 02138}

\altaffiltext{5}{Raman Research Institute, C.V. Raman Avenue, Bangalore, 560 080, India. Deceased 2001, October 29}

\begin{abstract}
We present high angular resolution ($0\rlap.{''}6$) observations 
made with the Very Large Array (VLA) of the radio continuum at 
8.3 and 43~GHz as well as  H92$\alpha$ and H53$\alpha$ radio 
recombination lines (RRLs) from the nearby ($\sim$3~Mpc) starburst
galaxy M82. In the continuum we identify 58 sources  at 8.3 GHz from 
which 19 have no counterparts in catalogs published at other frequencies.
At 43 GHz we identify 18 sources, unresolved at  $0\rlap.{''}6$ resolution, 
from which 5 were unknown previously.
The spatial distribution of the H92$\alpha$ line is inhomogeneous; 
we identify 27 features, about half of them are associated with continuum 
emission sources. Their sizes are typically in the range 2 to 10~pc.
Although observed with poorer signal to noise ratio, the H53$\alpha$ line is detected.
The line and continuum emission are modeled using a collection of HII regions 
at different distances from the nucleus. The observations can be interpreted 
assuming a single-density component but equally well with two components, if 
constraints originating from previous high-resolution 
continuum observations are used. 
The high-density component has a density of
$\sim 4 \times 10^{4}$~cm$^{-3}$. 
However, the bulk of the ionization is in regions with 
densities which are typically a factor 10 lower.

The gas kinematics, using the H92$\alpha$ 
line, confirms the presence of steep velocity
gradient (26~km~s$^{-1}$~arcsec$^{-1}$) in the 
nuclear region as previously reported, 
in particular from observations of 
the [Ne~II] line at $12$~$\micron$.
This gradient has about the same
amplitude on both sides of the nucleus.
As this steep gradient is observed not only on
the major axis but also at large distances along
a band of PA of $\sim 150^\circ$, the interpretation 
in terms of x2 orbits elongated along the minor 
axis of the bar, which would be observed at an angle 
close to the inclination of the main disk, seems inadequate. 
The observed kinematics cannot be modeled using a 
simple model that consists of a set of circular orbits 
observed at different tilt angles.
{\it Ad-hoc} radial motions must be introduced 
to reproduce the pattern of the velocity field. 
Different families of orbits are indicated as we 
detect a signature in the kinematics at the transition
between the two plateaus observed in the NIR light 
distribution. These H92$\alpha$ data also reveal the 
base of the outflow where the injection 
towards the halo on the Northern side occurs.
The outflow has a major effect on the observed kinematics, present
even in the disk at distances close to the nucleus.
The kinematical pattern suggests a connection between the gas 
flowing in the plane of M82 towards the center; this behavior 
most likely originates due to the presence of a bar and the 
outflow out of the plane.

\end{abstract}

\section{INTRODUCTION}

M82 is an excellent candidate to investigate the physical conditions 
of a starburst galaxy because it is one of the nearest ($\sim$3~Mpc) 
and brightest objects of this class. The bulk of the ionized gas in 
this galaxy is located in regions that are surrounded by large amounts 
of dust: in M82 the extinction in visual magnitudes, A$_{V}$, ranges 
from a few to about 15 mag.
Radio-wavelength observations, which are not affected by dust 
obscuration, can play a key role in the determination of the 
physical properties of the ionized gas in starburst galaxies.

M82 has been observed in the radio continuum over a wide range 
of frequencies (Kronberg, Biermann \& Schwab 1985; McDonald et 
al. 2002) and is known to host a population of compact sources 
as observed at angular resolutions $< 0\rlap{.}{''}$2 (McDonald 
et al. 2002 and references therein). The total number of compact 
sources decreases at 20~cm as compared to 2~cm; this result has 
been interpreted as due to free-free absorption by ionized gas in 
compact HII regions.
Of the 46 compact sources identified by McDonald et al. (2002), 
$\sim35\%$ were classified as HII regions based on their 
continuum spectra.

The distribution and kinematics of the ionized gas in the central kiloparsec 
of M82 have been previously studied using radio recombination lines (RRLs). 
The first detections of RRLs from M82 \citep{Se77, Ch77}
were a major achievement that led to further investigations of extragalactic RRLs.
VLA observations of RRLs up to 8~GHz by \citet{An90}
toward NGC 253 motivated further interferometric observations of RRLs at the same
frequency \citep{An93,Zh96,Mo02} and higher frequencies \citep{Zh00, An00} 
toward starburst galaxies.
Using the total integrated line emission, global estimates were made 
 for the properties of the ionized gas in M82 (Seaquist, Bell \& Bignell, 1985).
Using the Westerbork Synthesis Radio Telescope,
\citet{Ro87} obtained the velocity field at moderate angular resolution ($\sim 13''$) 
using the H166$\alpha$ RRL (at 1.4~GHz). 
These observations show the rotation of the ionized gas in the
central 600~pc, with solid body rotation within a radius of $\sim170$~pc or
$8''$. The velocity field has also been obtained using the [Ne~II] line
 in the mid IR with $2''$ angular resolution \citep{Ach95}. The velocity fields 
obtained from the H166$\alpha$ and the [Ne~II] line observations are not consistent,
specially in the SW half. However, the different angular resolutions achieved
for each line prevent a direct comparison. Higher angular
resolution observations of RRLs were necessary to understand the
kinematics of the ionized gas.
 \citet{Se96} observed the H41$\alpha$ line with angular resolution of 4$''$ 
and, in addition to the normal rotation, they showed the presence of kinematical 
features with velocity deviations up to 150~km~s$^{-1}$. From observations of 
the $^{12}$CO, $^{13}$CO and $^{18}$CO lines, Weiss et al.~(1999) and Matsushita 
et al.~(2000) reported evidence for an expanding supershell on the SW side of 
the nucleus of M82.

The high level of star formation activity at the center of M82 
could have been triggered due to the close interaction with the 
neighboring galaxy M81 and the presence of a bar that would
drive the gas inwards to feed the starburst. From observations of 
the morphology and the kinematics of the different constituents in the 
inner part of M82, the presence of x1 and x2 orbits has been suggested
to indicate the existence of a bar 
\citep{Ach95,Wi00,Gre02}. In this scenario, the ionized gas is 
mainly found along the x2 orbits i.e. highly confined near the center. 
X-ray observations also suggests the presence of 
a low luminosity AGN in M82
\citep{Ma99}.

The interpretation of the RRLs is not straightforward since 
the line emission mechanism could involve three different 
contributions: spontaneous as well as internal and external
stimulated emission.
The radio continuum emission in the starburst regions has 
contributions from free-free (thermal) and synchrotron 
(non-thermal) radiation, which also leads to complexity
in the interpretation of the observations. 
Using the radio continuum and RRL observations, \citet{An00} 
developed a model that consists of a collection of HII regions
in order to determine the physical properties of the ionized 
gas in Arp 220.
These authors have been able to reproduce the observations in 
the frequency range $0.15-113$~GHz and the simultaneous existence 
of both low density ($\sim$10$^{3}$~cm$^{-3}$) extended ($\sim$5~pc) 
HII regions and high-density ($\sim$10$^{5}$~cm$^{-3}$) ultra-compact 
($\sim$0.1~pc) HII regions was deduced.

In this paper, using the Vey Large Array (VLA) of the National Radio
Astronomy Observatory (NRAO), we present observations toward M82 of the 
radio continuum at 8.3 and 43~GHz and the H92$\alpha$ (3.6~cm) and 
H53$\alpha$ (7~mm) RRLs. 
With an angular resolution of $0\rlap.{''}6$ ($\sim$9~pc), we can obtain 
detailed information of the spatial distribution and kinematics of the 
ionized gas within the central starburst region.
In particular we search for continuum emission sources associated with 
the H92$\alpha$ and H53$\alpha$ line emitting regions.
This paper is organized as follows:
the observations are discussed in $\S$~2; the results are presented in 
$\S$~3; a model, based on the observations of radio continuum and RRLs, 
that consists of a conglomerate of HII regions is presented in $\S$~4; 
a discussion of the radio continuum emission at 8.3 and 43~GHz, the 
implications of the kinematics and the results of the proposed model 
are discussed in $\S$~5; and finally, the conclusions are summarized 
in $\S$~6.

\section{OBSERVATIONS.}

\subsection{8.3~GHz data.}
\subsubsection{Observations and calibration}
The observations of the H92$\alpha$ line ($\nu_{rest}=$8309.3832~MHz)
were conducted in the C (April 22, 1996), CnB (Feb 13, 2000) and B
(May 05 and 11, 2001) VLA configurations. The maximum angular resolution 
achieved is $\sim 0\rlap.{''}6$.
The observations made in the C array have been acquired from the VLA 
archive database.
The 31 spectral channels mode was used with a total bandwidth of 25~MHz,
corresponding to a velocity coverage of $\sim$850~km~s$^{-1}$,
centered at a heliocentric velocity of 200~km~s$^{-1}$.
The flux density scale was determined by observing 3C286
(5.3~Jy at 3.6~cm).
The phase calibrator was 1044+719, with a flux density of 1.5~Jy.
The bandpass calibration was made using the calibrator 3C48 with a 
flux density of 3.2~Jy. Bandpass calibration is critical in these 
observations because the line-to-continuum ratios are $\le 1$\%.
The data were Hanning-smoothed offline to improve the signal-to-noise
ratio (S/N) and minimize the Gibbs effect.
The effective velocity resolution is 56~km~s$^{-1}$ (806~kHz).
Each of the databases were self-calibrated in phase using the continuum
channel; the solutions were then applied to the spectral line data.
The data sets taken at different epochs were combined into a single data set.
The self-calibration process was further used on this data set to correct
for small phase offsets between the individual observations.
The AIPS task UVLSF was used to estimate the continuum level by
fitting a linear function through the spectral channels
free of line emission, and then subtracted from each visibility
record. The observational parameters for the H92$\alpha$ line are listed in Table~1.

\subsubsection{Imaging.} 
The use of a natural weighting scheme for the data in the C configuration relative 
to those in the B configuration produces a synthesized  beam with prominent wings 
(at the level of 2\%). 
In order to correct this problem, we use a technique that consists of re-weighting 
the data in the {\it u,v} plane so as to produce a gaussian beam in the image.
After the re-weighting process, if necessary, the image 
is deconvolved using the CLEAN algorithm. 
Because the {\it u,v} plane is properly sampled in the inner region, the  
deconvolution is not required when producing images at an angular resolution of $2''$ FWHM. 
For the images made at angular resolutions of $0\rlap.{''}6$ and $0\rlap.{''}9$
(section 3), a deconvolution is required. 
In these cases the $2''$ angular resolution image is used as an input for
regularization in the process of deconvolution. The quality of the final images 
is limited by errors due to the imperfect continuum subtraction. These 
errors appear as weak fluctuations over a scale of one arcmin.
These fluctuations have a typical correlation length of about 4~MHz along the 
spectral axis. This behavior results in systematic errors which limit the quality 
of the ``baselines'' ($\sim 2\%$) in the spectra produced from different regions of M82
for our data analysis.

\subsection{43~GHz data}
\subsubsection{Observations and calibration}
The observations of the H53$\alpha$ line ($\nu_{rest}=$42951.9714~MHz) were
carried out in the C array of the VLA on April 13, 2000.
The VLA correlator is limited to a bandwidth of 50 MHz ($\sim$350~km~s$^{-1}$ at 7~mm).
At 43 GHz three adjacent spectral windows, each of them with 15 channels, 
are required to cover the velocity range of the RRL ($\sim$600 km s$^{-1}$).
The windows were centered at 42185.1, 42214.9 and 42235.1 MHz.
The amplitude, phase and bandpass calibrators used were 3C286 (1.47~Jy),
0954+658 (0.87~Jy) and 1226+023 (19.4~Jy), respectively.
The flux density calibrator was observed only for
the central LO window (42214.9~MHz). The flux densities of
the phase and bandpass calibrators in the adjacent LO windows (42185.1 and 42235.1~MHz)
were assumed to have the same values as those determined from the central LO window.
The phase calibrator was observed at time intervals of 10~min.
The on-source integration time was $\sim$2~hrs for each frequency window.
The bandpass response of the instrument is different for each
frequency window; the observation of both the bandpass and the
phase calibrator must be interleaved between the different
overlapping frequency windows required to observe the complete line.
This method removes the offsets ($\le 5\%$) between the frequency windows.
In order to correct for phase decorrelation that may be present for time intervals $<$8~min,
a phase calibration was performed initially followed by a second calibration step
applied to both phase and amplitude. Phase correction is more important at these
frequencies since the troposphere introduces phase offsets that may affect the 
coherence of the data. 
The line-to-continuum ratio for this line is  $10-20\%$ in the brightest region of M82.
Table 1 shows the details of the observations for the H53$\alpha$ line.

\subsubsection{Imaging.} 

In order to produce the H53$\alpha$ line image, 
the contribution of the continuum emission must be
removed and the three frequency windows combined. 
The method uses two {\it a priori}  constraints
on the source model that are not independent. 
The first constraint is set by assuming that the emission in 
the lowest spatial frequencies is mostly continuum emission, 
allowing us to clip the data for the shortest spacings in the 
{\it u,v} plane. 
The second constraint is based on the fact that the projected 
velocity of the gas in M82 varies across 
the image due to the rotation of the galaxy. 
Hence, if we include {\it a priori} information on the rotational 
properties of M82, for each position in the image, we can maximize 
the number of line-free channels in each LO setting to reduce the 
uncertainties when removing this contribution. 
As a result this second method involves the following steps for 
each LO setting:
(1) filter out part of the emission present at the lowest spatial 
frequencies,
(2) produce an undeconvolved 3-dimensional image,
(3) determine the level of remaining continuum for each position,
(4) determine the offsets by comparing these levels from one LO setting 
to the other for each position,
(5) refine the determination of this remaining continuum by including 
the line-free channels of the adjacent LO settings for each position,
(6) subtract the determined continuum contribution and
(7) average the three spectral windows to make the single
3-dimensional line image.
It is not necessary to deconvolve the results after step 6 
because the signal-to-noise ratio for the line emission is between 2 and 5.

In order to produce the 43 GHz continuum image we carried out the following 
steps:
(1) produce a raw 3-dimensional image for each LO setting, 
(2) subtract the line contribution as derived above in each LO setting,
(3) determine the mean continuum level for each position in each LO
setting,
(4) average the three 2-dimensional images obtained in step 3 and 
also average the three synthesized beams and
(5) finally deconvolve this averaged image.

The undeconvolved images were generated using the AIPS IMAGR program 
applying the suitable {\it u,v} tapers to obtain images at different angular 
resolutions.
They were further processed in the GIPSY environment to
perform all the subsequent steps. The data were Hanning-smoothed offline
and a final spectral resolution of 44~km~s$^{-1}$ was achieved. 

\section{RESULTS}

Globally, from the 8.3~GHz data, we obtain a 
total integrated continuum flux density of 
2.6$\pm$0.1~Jy and a H92$\alpha$ line flux 
density of 2.9$\pm$0.1~Jy~km~s$^{-1}$. At 
43~GHz the total continuum flux density  
is in this case 0.82$\pm$0.16~Jy and the H53$\alpha$ line 
flux density is 3.0$\pm$0.2~Jy~km~s$^{-1}$.
Images at three different resolutions are produced to analyze the data. 
The high angular resolution ($0\rlap.{''}6$) images are suitable to investigate the compact bright features. 
Using the intermediate resolution images ($0\rlap.{''}9$), it is possible to study the weak extended features 
with sufficient angular resolution.
The low angular resolution images ($2''$)  
provide information on the overall distribution of the ionized gas.

\subsection{Radio continuum at 8.3 and 43~GHz}

\begin{figure}[!ht]
\plotone{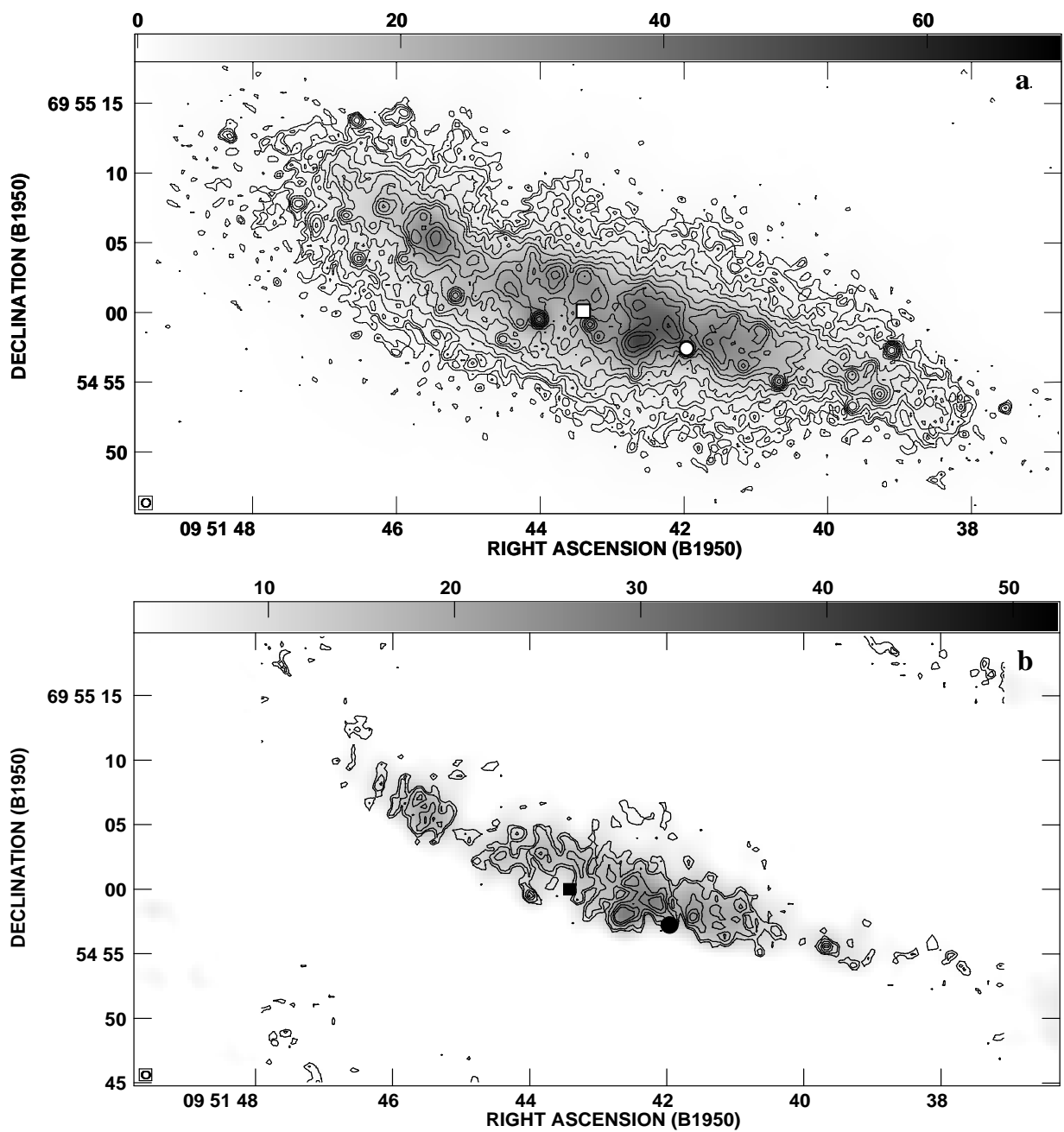}
\caption{
a) Radio continuum images at 8.3~GHz made with the VLA in the B and C arrays.
Contour levels are drawn at -3, 3, ..., 269 times the rms noise of 
0.07~mJy~beam$^{-1}$ at steps of 2$^{1/2}$. 
The gray scale covers the range from 0 to 70~mJy~beam$^{-1}$. 
b) Radio continuum images at 43~GHz made in the C array of the VLA.
Contour levels are -3, 3, 4, 6, 8, 10 times 0.4~mJy~beam$^{-1}$, 
the rms noise in the central regions. 
The gray scale image covers the range from 3~mJy~beam$^{-1}$ to 53~mJy~beam$^{-1}$.
In both images, the radio continuum made with high angular 
resolution of $0\rlap.{''}6$ (contours) is superposed on the 
radio continuum image with low angular resolution of $2''$ 
(gray scale). The square shows the position of the 2.2~$\micron$ 
peak \citep{Les90} and the circle marks the position of the SNR G41.95+57.5. 
}
\label{figcont}
\end{figure}

A correction for the primary beam was applied to the images at 8.3~GHz and 43~GHz.
In the images at 8.3~GHz this correction 
does not exceed  $\sim$2\%, while for the  43~GHz images this 
correction is a factor of 1.2  at the extreme edges of M82.
The radio continuum images at 8.3 and 43~GHz are shown in 
Fig.~\ref{figcont}a and Fig.~\ref{figcont}b, respectively;
the high angular resolution images (contours) are shown overlaid on 
their corresponding low angular resolution images (gray scale) at each 
frequency.

\begin{figure}[!ht]
\plotone{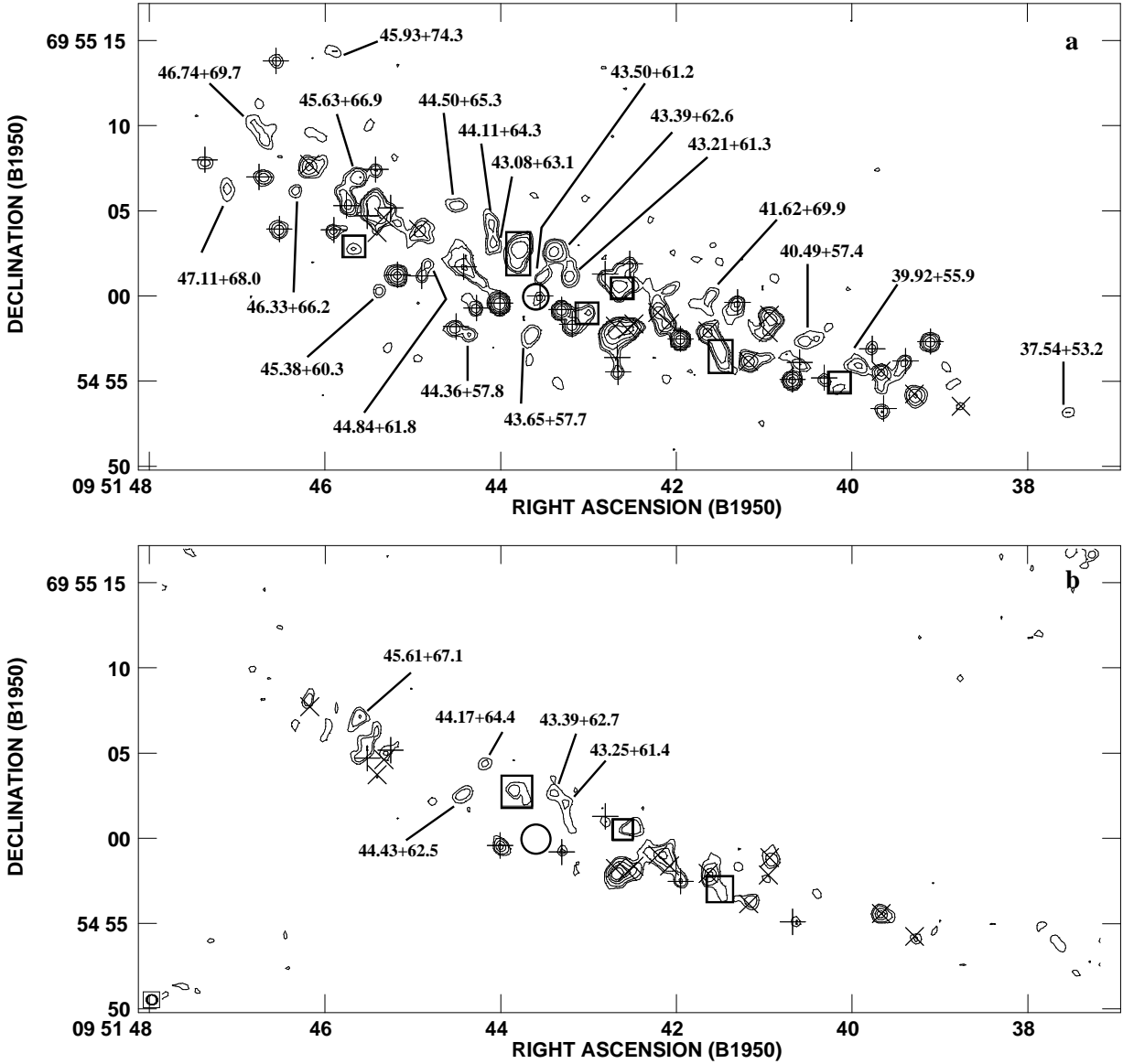}
\caption{Radio continuum images after subtracting the background local emission.
(a) 8.3~GHz image (B and C arrays).
Contour levels are -3, 3, 6, 12, 24, 48, 96 times the rms noise of
0.04~mJy~beam$^{-1}$.
(b) 43~GHz image (C array).
Contour levels are drawn at -3, 3, 4, 6, 8, 10 times the rms noise of 0.4~mJy~beam$^{-1}$.
Both images have angular resolution of $0\rlap.{''}6$.
The crosses (+) indicate the position of SNRs, while the rotated crosses (X) 
indicate the HII regions \citep{Mc02}.
The empty squares show the position of sources previously observed, for which
the spectral index  has not been determined. The empty circle marks the position of
the 2.2~$\mu$m peak \citep{Les90}.
The newly detected features  are labeled according to their positions (e.g. 39.92+55.9).}
\label{figcompcont}
\end{figure}

A number of previous high angular resolution ($\le 0.1''$) studies 
carried out over the range 408~MHz to 15~GHz have revealed the existence 
of a population of compact sources classified as HII regions or 
SNRs \citep{Kr85,Hu94,Mu94,Wi97,All98,Mc02}.
In order to study the small scale continuum features at 8.3~GHz, 
the underlying extended emission was filtered out by restricting 
the lower uv-baseline range (to {\it u,v} distances larger than 
50~k$\lambda$) so that no structures larger than 4$''$ are present. 
At 43~GHz an iterative procedure is used to determine and remove the 
spatially extended continuum emission.
Figures \ref{figcompcont}a and \ref{figcompcont}b show the resulting 
spatial distribution of these features at 8.3 and 43~GHz, respectively.
In these images we are able to identify  $\sim 85 \%$ of the previously 
known compact sources.
The remaining $15\%$ of compact sources were not detected because
their peak flux density is below the level of the local diffuse emission.
On the other hand, these images reveal 21 new features (19 of these 
sources are detected at both frequencies and two sources only at 43~GHz) 
not detected in previous studies.
Several of these sources are relatively more extended ($> 0\rlap.{''}5$) 
and they could not be identified in the previous high angular resolution images.
Including our new detections, a total of 60 sources are detected toward M82, 
58 features at 8.3~GHz and 18 at 43~GHz.

\begin{figure}[!ht]
\plotone{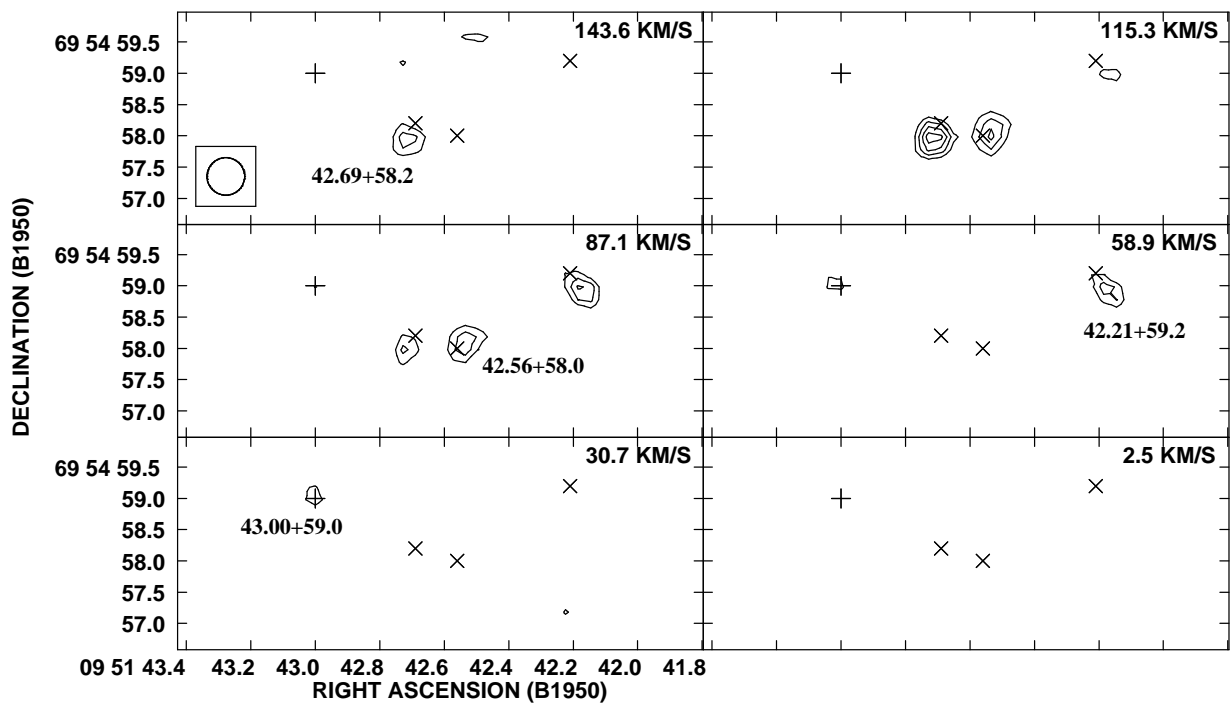}
\caption{High angular resolution images of the H92$\alpha$ line emission (contours) 
throughout the velocity region where emission was observed in the central region of M82. 
The beam size is $0\rlap.{''}6$, shown in the bottom left corner of the first image.
The contour levels are -3, 3, 4, 5, 6  times the rms noise of $\sim$0.12~mJy~beam$^{-1}$. 
The central heliocentric velocity is given above each image. 
The cross (+) indicates the position of the compact source 43.00+59.0, classified in this work
as SNR (see discussion in section 5), and the rotated crosses (X) indicate 
the HII regions; all of these are compact sources observed
in the continuum \citep{Mc02,Hu94} and coincide with features observed in the H92$\alpha$ line. }
\label{figh92hr}
\end{figure}

In Table~2 we list the parameters of these 60 compact features:
\begin{itemize}
\item The first column lists the galactic names for each source.
For each feature we measure its spatial position, its deconvolved 
angular size and flux density. 
The measurements at 8.3~GHz are summarized in columns 2, 3 and 4 
and those at 43~GHz in columns 5, 6 and 7, respectively. 
For the 21 newly detected features a super-script ``a'' is used.
The uncertainties in flux densities correspond to a 1 
$\sigma_{rms}$ noise in the least-squares fitting procedure. 
For features not detected, we provide an upper limit of 
3 $\sigma_{rms}$ for the flux densities.
\item From the measured flux densities (columns 4 and 7)
we derive the spectral index (column 8) between 8.3 and 43~GHz. 
This value can be compared with the spectral index previously reported in the 
literature (column 9) using different frequency ranges (e.g. 0.408 to 15~GHz). 
\item In column 10 we list identifications for the sources which have 
radio recombination line emission. The associations are made with the 
H92$\alpha$ line features observed at 0$\rlap{.}{''}$9 (see section 3.2). 
At an angular resolution of 0$\rlap{.}{''}$6, only four H92$\alpha$ compact line features
are detected with signal-to-noise ratio $> 3$ (see section 3.2 
and Fig \ref{figh92hr}).
\item In the last column, the classification is given
(HII regions or SNR).
The sources previously identified as HII regions by \citet{Mc02} were 
classified based on their spectral indexes, $\alpha$, which in these 
cases are inverted ($S_{\nu} \propto \nu^{\alpha}$, $\alpha>0$) 
between 5 and 15~GHz (column 9). 
We confirm the classification for all the HII regions previously reported,
as they have a spectral index that is consistent with optically thin 
free-free emission between 8.3 and 43~GHz (column 8). 
We assign tentatively an HII type to four more features, because between  
8.3 and 43~GHz, 43.21+61.3 and 45.63+66.9 have flat spectra and
44.17+64.4 and 44.43+62.5 have inverted spectra.
For the sources previously identified as SNRs, our results show
non-thermal spectra between 8.3 and 43~GHz (see column 8) as expected 
from the spectral index computed between 5 and 15~GHz.
In addition to those features previously classified as SNRs, based on 
the upper limits obtained from the spectral index between 8.3 and
43~GHz, we can tentatively consider 39.92+55.9, 40.49+57.4,
43.00+59.0, 44.11+64.3 and 46.74+69.7 as SNRs. 
It is noted that three SNRs, 41.29+59.7, 43.00+59.0 and 45.79+65.2, 
exhibit H92$\alpha$ RRL emission; this issue will be discussed in $\S$~5.
\end{itemize}

\subsection{Radio Recombination Lines H92$\alpha$ and H53$\alpha$}

Figure~\ref{figh92hr} shows the  high-resolution  ($0\rlap.{''}6$) 
H92$\alpha$ line images in the region where H92$\alpha$ RRL emission 
is detected; in these images four sources are
clearly identified.
The center of this region is located $\sim$4$''$ W of the 2.2~$\mu$m peak 
\citep{Les90} and 
corresponds to the brightest area observed in the H92$\alpha$ line
in M82 on a larger scale.
All four line emitting sources are spatially associated with
compact continuum sources previously identified by 
\citet{Mc02}. No other line features are detected outside this 
region at this resolution.
For Figure~\ref{figh92ir} we show the velocity-channels 
at an intermediate angular resolution of $0\rlap.{''}9$ (top) and an image of
the total integrated H92$\alpha$ line emission (bottom).
In this case  27 line features can be identified and those associated 
with continuum sources are listed in column~(10) of Table~2.
In these images,  the position of the HII regions (crosses) and SNRs 
(stars) are shown as reported by \citet{Mc02}. 
In this comparison the SNRs are mainly located at the periphery of 
the line emitting regions while the HII regions are clearly associated 
with the RRL emission.
Because these HII regions are the compact continuum sources which have 
inverted spectra between 5 and 15~GHz (section 3), 
we have direct evidence that this population of compact 
objects is associated with the more extended HII regions which 
produce the H92$\alpha$ recombination line emission.
Figure~\ref{figh92lr} (top) shows the H92$\alpha$ line velocity-channels 
 at a low angular resolution of $2''$ (top) for the overall
distribution of the integrated H92$\alpha$ line emitting regions (bottom). 

Figures \ref{figlandc}a and \ref{figlandc}b show the low-resolution
images of the integrated H92$\alpha$ and H53$\alpha$ line emission 
(contours) overlaid on the low-resolution images of radio continuum 
at 8.3 and 43~GHz (gray scale), respectively. 
To produce these images for the integrated line emission only the 
regions with line emission $ > 3 \sigma_{rms}$ were selected. 
There is a clear correspondence between the spatial distribution of the 
line and the continuum emissions. In both cases the distribution of the 
total line emission is characterized by a pair of concentrations on
each side of the center defined by the 2.2~$\mu$m peak \citep{Les90}. 
However, the sources located on the E side are fainter compared 
with their counterpart on the W side. 
This asymmetry has also been observed in [Ne~II] and for several molecules 
\citep{Ach95}.
Figure~\ref{figlandc}c shows the velocity integrated line emission of 
the H53$\alpha$ line (contours) overlaid on the corresponding image of 
the H92$\alpha$ line (gray scale).
A good correspondence between the spatial distributions for the two 
lines is observed. Both lines cover an angular size of about $30''$ 
(480 pc) along the major axis and the main peak is $\sim 4 ''$ W of 
the 2.2~$\mu$m peak.
However, we note that the H92$\alpha$ line emission may be more
extended than the H53$\alpha$ line emission ($\sim$ 4$''$) on the
extreme NE side of M82. 
More sensitive H53$\alpha$ line observations would be required to
verify this possible difference. 
In order to characterize the physical properties of the ionized gas in 
M82 (section 4.2), we define five regions ( E1, E2, C, W1 and W2) as 
indicated by the rectangular boxes in Figure~\ref{figlandc}c. Regions
E1, W1 and W2 have been labeled according to peaks observed in
[Ne~II] emission \citep{Ach95}. 
The comparison of the distribution of the H92$\alpha$ line emission with 
the distribution of other tracers of the ionized gas is discussed in $\S$~5.2.

Figure \ref{figspec} shows the spatially integrated H92$\alpha$ and 
H53$\alpha$ spectra for four distinct regions ( E1, E2, W1 and W2); 
the spectrum for region ``C'' is not shown since H53$\alpha$ line emission 
is not detected above 3$\sigma_{rms}$.
The area for the integration of line and continuum emission
are defined by the first contour (57~Jy~beam$^{-1}$~m~s$^{-1}$) of the 
H53$\alpha$ distribution.
The H92$\alpha$ and H53$\alpha$ line parameters derived from the fits 
of gaussian profiles (Figure \ref{figspec}) are given in Table~3. 
Columns (2), (3) and (4) of Table~3 list the parameters obtained for 
the RRL H92$\alpha$ and column (5) the continuum flux density at 
8.3~GHz. Columns (6), (7) and (8) list the parameters obtained for the
RRL H53$\alpha$, column (9) the continuum flux density at
43~GHz, column (10) the maximum filling factor for the
high-density component explained in $\S$~4 and column (11)
the continuum flux density at 5~GHz (personal communication, A. Pedlar). 
These measurements form the basis for the models discussed in $\S$~4.

The global properties for the five regions identified in 
Figure~\ref{figlandc}c are given in Table 4, the total H92$\alpha$ 
line emission integrated over each rectangular box (column 2) and 
the corresponding line (H92$\alpha$) to continuum (at 8.3~GHz) ratio 
(column 3). Inside these four regions a number of sub-regions have 
been identified in the H92$\alpha$ line image (see Table~4). 
For each of these sub-regions (labeled in column 4), Table~4 lists
the spatial coordinates (column 5), the velocity-integrated line emission
(column 6), the FWHM angular size deconvolved to correct for the $0\rlap.{''}9$ 
beam broadening (column 7), the peak flux density (column 8), 
centroid velocity (column 9) and the spectral width (column 10). 
The angular size of each sub-region is based on fits of 2D gaussians 
in the appropriate spectral channels; the spectral line parameters were 
obtained fitting a gaussian profile along the spectral axis for the spatially 
integrated emission in each of these sub-regions. For each region the sum of 
the line flux densities of the sub-regions is also given and can be compared with the 
total line flux density in column 2, determined from the low angular resolution 
image. For all regions, essentially all the line flux density 
arises from these sub-regions. Thus, any diffuse emission is $\le 10\%$.

The velocity field determined using the H92$\alpha$ line 
image at 2$''$ angular resolution is shown in Figure~\ref{figvf}a. 
In this image, the major and minor axis are indicated for a disk 
with a diameter of 40$''$ as observed at an inclination of $81^\circ$, 
and position angle (PA) of $68^\circ$ relative to the observer. 
The intersection of these two axes is set at the position of the 
2.2~$\micron$ peak \citep{Di86}, which also coincides with 
the center of the hole in Fig.~5b. 
The position of the center of the ring fitted by \citet{Ach95}, 
assumed to be the kinematic center, is indicated by a cross. 
The important features in this velocity field are the 
following: 
{\it a)} The major axis corresponds to the line of nodes 
which gives the maximum velocity gradient for the outer regions,
{\it b)} in the central region there is a significant tilt of 
the iso-velocities relative to the PA of the minor axis,
{\it c)} this velocity field is dominated by an axisymmetric 
pattern and the position of the center of symmetry appears to 
have a small offset ($\sim 1\rlap.{''}5$) relative to the position 
of the 2.2~$\micron$ peak,
{\it d)} Some areas show significant deviations relative to this 
general pattern, in particular in the region near 
9$^h$51$^m$43$\rlap.{^{s}}$5,~+69$^{\circ}$55$'$04$''$, where the 
velocities are $\sim 40$~km~s$^{-1}$ larger than expected. 
The kinematics of the ionized gas are discussed in $\S$~5.3.
The spatial distribution of the velocity dispersion is shown 
in Figure \ref{figvf}b.
One important feature in this image is the presence of a 
band $\sim 1\rlap.{''}5$ wide with a PA of $\sim 150^\circ$,
close to the minor axis, where the spectra are systematically narrower. 
This feature is analyzed in section 5.3.

\section{Models with multiple HII regions.}

In three starburst galaxies (Arp 220, M83, 
NGC 2146) a model consisting of a collection 
of HII regions has been used successfully to 
explain the observations of RRLs and the 
radio continuum \citep{An93, Zh96, An96, An00}.
The ionized gas distribution in the center 
of M82 is inhomogeneous as observed in the 
RRLs H92$\alpha$ and H53$\alpha$.
Thus, a model that consists of a collection
of HII regions seems appropriate to explain 
the RRL and radio continuum emission from this galaxy.
Based on this model we can derive the physical 
parameters of the HII regions (e.g. electron
temperature and volume density).
The model must reproduce both the observations of the 
line and continuum emission as a function 
of frequency. For M82, observations at 
2$''$ angular resolution of RRLs are available
 at two different frequencies (8.3 and 43~GHz) 
and observations of radio continuum at three 
frequencies (5, 8.3 and 43~GHz).
The 5 GHz flux densities are listed in column~11 of Table~3.

The gas is ionized by young massive stars (types O and/or 
B) radiating large amounts ($\ge 10^{49}$~s$^{-1}$) of 
Lyman continuum photons.
The number of Lyman continuum photons determines the 
size, $l$, of each HII region depending on the electron 
density. 
Since these models are constructed to obtain average values for 
the physical properties of the ionized gas, we have assumed: 
(1) Each ionizing massive star emits the same number of Lyman 
continuum photons (N$_{Lyc}=10^{49}$~$s^{-1}$) and 
(2) N$_{Lyc}$ is related to the electron density and the size 
of each HII region by  $N_{Lyc} \propto n_{e}^{2} l^{3}$; this 
relation implies that for each density value, $l$ will be 
determined and then $l$ is not a free parameter in the model, 
providing a constraint.
Since massive stars have a relatively short lifetime on the 
main sequence, the derived number of Lyman continuum photons 
can be related to the star formation rate (SFR) of O and B 
stars if  the absorption of Lyman continuum photons by dust 
inside the HII regions is neglected.

This model consists of HII regions located in front of a diffuse 
mixture of thermal and non-thermal emission (S$_{Cbg}$).
The HII regions radiate free-free emission and recombination lines.
Background non-thermal emission and free-free emission
that arise inside each HII region may stimulate the emission of 
RRLs in the HII regions.
Thus, the line emission from each HII region has three different 
contributions:
(i) spontaneous, (ii) internally stimulated and (iii) externally 
stimulated emission. 
Each HII region is characterized by an electron temperature $T_{e}$~(K), 
electron density $n_{e}$~(cm$^{-3}$), and linear size $l$ (pc). 
The thermal continuum flux density, S$_{C-TH}$, is given by,

\begin{equation}\label{contflux}
S_{C-TH}=\Omega_{HII} B_{\nu}(1-e^{-\tau_{c}})~mJy,
\end{equation}

\noindent where $\Omega_{HII}$ is the solid angle subtended by 
each HII region, $B_{\nu}$ is the black-body emission (mJy) and $\tau_{C}$ 
is the continuum optical depth, which depends on frequency $\nu_{L}$ 
as follows \citep{Be78}:

\begin{equation}\label{tauc}
\tau_{C}=3.01\times 10^{-2} \nu^{-2} T_{e}^{-1.5}ln 
\bigg( \frac{4.95 \times 10^{-2} T_{e}^{1.5}}{\nu_{L}} \bigg) EM_{C},
\end{equation}

\noindent where $EM_{C}=n_{e}^{2}l$ is the emission measure for 
each HII region. The line flux density, $S_{L}$~(mJy), arising 
from each HII region is calculated using,

\begin{equation}\label{lineflux}
S_{L}=\frac{2k\nu^{2}}{c^{2}}\Omega_{HII}T_{e}
\biggl[\frac{b_{n}\tau_{L}^{*}+\tau_{c}}{\tau_{L}+\tau_{c}}
(1-e^{-\tau_{L}+\tau_{c}})-(1-e^{\tau_{c}})\biggr]+S_{Cbg}
e^{-\tau_{c}}(e^{-\tau_{L}}-1)~mJy.
\end{equation}

The spontaneous and the internally stimulated emission are represented 
by the first term in Eq. \ref{lineflux} and the externally stimulated 
emission is represented by the second term.
$k$ is the Boltzmann constant, c is the speed of light, 
$\tau_{L}^{*}$ is the peak line optical depth under LTE conditions, 
$\tau_{L}=b_{n}\beta_{n}\tau_{L}^{*}$ is the peak line optical depth 
corrected for non-LTE effects, $b_{n}$ is the departure coefficient, 
$n$ is the quantum number, and $\beta_{n}=1-(k T_{e}/h \nu_{L})d ln b_{n}/dn$. 
For a given combination of parameters T$_{e}$ and n$_{e}$, we compute
the corresponding values of $b_{n}$ and $\beta_{n}$ \citep{Sal79} which 
are used to calculate the integrated line flux density of a single HII 
region using Eq.~\ref{lineflux}.
The line optical depth is given by $\tau_{L}^{*}=\tau_{c}S_{L}/S_{C-TH}$ 
where the line to thermal continuum ratio in LTE is defined as,

\begin{equation}\label{linetocont}
\frac{S_{L}}{S_{C-TH}}=2.33 \times 10^{4} \Big( \frac 
{\Delta \nu_{L}}{kHz}\Big)^{-1}
 \Big( \frac {\nu_{L}}{GHz} \Big)^{-2.1} \Big( \frac 
{T_{e}}{K} \Big)^{-1.15}
  \frac {EM_{L}}{EM_{C}},
\end{equation}

\noindent where EM$_{L}$ is the emission measure from hydrogen 
and EM$_{C}$ is the emission measure from all ions (cm$^{-6}$~pc).
In these models, EM$_{L}$/EM$_{C}=1$; for simplicity 
we do not consider the presence of ionized He.

Single-density (SD) models, consisting of a number of HII regions with
the same physical properties, have been used to reproduce the observations.
Additionally, following \citet{An00}, two-density (TD) models that consist of both low- and 
high-density HII regions were used to reproduce the observed 
values (listed in Table~4) for four regions ( E1, E2, W1, W2); 
in region 'C' the H53$\alpha$ line emission is not 
detected (see Figure~\ref{figlandc}c). 
The following constraints have been imposed: (i) the predicted 
thermal radio continuum flux densities should not exceed the radio 
continuum flux density observed at 43~GHz, 
(ii) the area filling factor should be $\le 1$,
(iii) the peak line emission from the model should not be larger 
than the corresponding observations,  
(iv) the spectral index of the background radiation has been 
limited to be not steeper than $-$1.5,
(v) the continuum computed from the model must reproduce all 
continuum observations at 6~cm, 3.6~cm and 7~mm, and
(vi) the area filling factor for the high-density component has an upper
limit imposed by observations of compact continuum sources at high
angular resolution \citep{Mc02}. 
For the two-density models the constraint (vi) is based on the 
additional assumption that the high-density HII regions are 
coincident with compact sources previously identified as HII regions.
Using the emission measure given in Table 4 of \citet{Mc02}, we compute 
the angular sizes and the maximum filling factor for the HII regions 
that form the high-density component.
The models that are consistent with these constraints were accepted 
as valid solutions.

The number of HII regions for the single-density model is computed by 
dividing the observed integrated line flux density by the integrated 
line flux density from a single HII region.
In the case of the model with two-density components, the number of HII 
regions with high-density gas (N$_{HII-HD}$) is computed dividing 
the area filling factor derived above by the area filling factor that a 
single high-density HII region can occupy (following $n_{e}\propto l^{-3/2}$).
Once we know the contribution of the high-density component to the line 
and continuum emission, the number of HII regions in the low-density 
component (N$_{HII-LD}$) can be estimated (see Table~5)  .

For both the single-density and the two-density models,
the contribution from the background emission is computed as the difference
between the observed radio continuum at 8.3~GHz and the value predicted by 
the models for the thermal emission at this frequency.
The spectral index value ($\alpha$), determined for this background emission, 
is obtained by minimizing the difference between the observed and the predicted 
values of the thermal continuum flux density at 43~GHz.

Since we have assumed that each HII region is being excited by stars that 
emits N$_{Lyc}=10^{49}$ photons per second, the total rate of emission 
of ionizing photons, N$_{Lyc-tot}$,  is,

\begin{equation}\label{nlyctot1}
N_{Lyc-tot}=N_{Lyc}N_{HII},
\end{equation}

\noindent
where $N_{HII}=N_{HII-LD}+N_{HII-HD}$. Another way to estimate N$_{Lyc-tot}$ 
is using \citep{Sch69,Ro80},

\begin{equation}\label{nlyctot2}
\bigg[ \frac{N_{Lyc-tot}}{s^{-1}} \bigg] = 9.0 \times 10^{43}\bigg( 
\frac{S_{C-TH}}{mJy}\bigg) \bigg( \frac{Te}{10^{4}~K}  \bigg)^{0.35}
\bigg( \frac{\nu}{4.9~GHz} \bigg)^{0.1} \bigg( \frac{D}{kpc} \bigg)^{2},
\end{equation}

\noindent where we can use the flux density from thermal emission, $S_{C-TH}$, 
at any frequency (if $\tau_{C}<<1$)  and D, the distance to the galaxy. 
In the case of optically thin emission at 43 GHz, both estimates of
N$_{Lyc}$ give similar values. However, the model is based on a 
number of assumptions. On the other hand, the number of Lyman continuum
photons is proportional to the radio continuum at 43 GHz and does not
depend on extinction estimation as occurs for the near-infrared case. 
Thus, the determination of the number of Lyman continuum photons 
from the continuum at 43 GHz is the most reliable way to estimate 
the Lyman continuum photons production.
If the gas is optically thick (which may be the case for a high-density component) 
at 43 GHz, then the estimation of the total number of ionizing photons using 
Eq. \ref{nlyctot1} is a lower limit.
The total SFR can be derived from the Lyman continuum luminosity following
\citet{An00}.

\subsection{Results from the models.}

The parameters T$_{e}$ and n$_{e}$ were considered to lie in 
the ranges $5000-10,000$~K (Garay \& Rodr\'{\i}guez 1983) and 
500-10$^{6}$~cm$^{-3}$, respectively.
The low-density component was defined to consist of HII regions 
with electron density, n$_{e}$ in the range 100$-$10$^{4}$~cm$^{-3}$, 
while the high-density component was defined as composed of HII regions with  n$_{e}$ 
in the range $2 \times 10^{4}-10^{6}$~cm$^{-3}$.
In Table~5 we list the range of physical parameters estimated 
using the models for each identified complex of HII regions 
(see Fig.~\ref{figlandc}c).
For each identified complex of HII regions, Table~5 lists: 
the size of each HII region, number of HII regions, factor of area 
covered by each density component, number of ionizing photons, 
star formation rate, line and continuum opacity at 8.3 GHz,
the departure coefficients for the H92$\alpha$ line, the spectral 
index of the background emission and the total mass of ionized gas.
Figure~\ref{figmods} shows the expected variation of radio continuum 
and the integrated RRL strength as a function of frequency for regions 
E1, E2, W1 and W2. The models shown in Figure~\ref{figmods} correspond 
to the mean values based on the models listed in Table~5.

The single-density models reproduce the observed 
line and continuum flux densities for all regions considered.
However, these results only provide a rough estimate 
for the physical properties of the ionized gas. As we know 
from previous studies \citep{Mc02}, there are compact 
HII regions in which higher density gas ($> 10^{3}$~cm$^{-3}$) could be present.
Using the area filling factor constraint in the two-density models 
(see section 4), we obtain valid solutions for regions E2, 
W1 and W2. These two densities are a few $10^{3}$~cm$^{-3}$ and 
 $10^{4}$~cm$^{-3}$ as listed in Table~5. 
For the region E1, there is no evidence of 
any compact high-density HII regions \citep{Mc02}.
Since for region E1 we do not have an area filling factor constraint, 
an estimate of the physical properties of the high-density ionized gas was not computed.

Based on these models the area filling factor from all HII regions, 
including the cases where the high-density component is taken into 
account, is $\sim 0.3$. 
We consider that the number of HII regions (obtained from the models)
 in the compact features is an upper limit.
As listed in Table 5, the low-density ionized gas has 
an electron density of $\sim 5 \times 10^{3}$~cm$^{-3}$.
\citet{Mc02} found that the turnover in the 
spectrum of SNRs at 1.4~GHz requires foreground emission measures 
of $\sim 10^7$ pc cm$^{-6}$, consistent with our model.  
However, it is likely that there is also ionized gas in a diffuse 
component with lower electron densities. 
\citet{Mc95} estimated the electron density to be typically  
$1.5 \times 10^{3}$~cm$^{-3}$ in the plane of M82 near the center of the galaxy. 
In order to explain observations in the far infrared
(made at an angular resolution of 80$''$), \citet{Co99} presented a model which 
is a combination of HII regions with $n_{e} \sim 0.25 \times 10^{3}$~cm$^{-3}$ 
and photo-dissociated regions (PDRs) with $n_{e} \sim 2 \times 10^{3}$~cm$^{-3}$. 
High resolution observations of RRLs of higher quantum number (e.g. H116$\alpha$), 
with an angular resolution similar to the H92$\alpha$ data presented here, 
are required to constrain the models to account for the gas at lower density.

Assuming that the relative number of high-density HII 
regions compared to the number of low-density HII regions is 
an age indicator of recent star formation activity, based on
the two-density models, we infer that the region W2 is the youngest. 
Another age indicator for the starburst in each region is the spectral 
index $\alpha$: the steeper the spectral 
index the more evolved the star formation activity, which is true
if we assume that each region has formed stars at a constant rate, 
at least during the last few $10^{6}$~yr.
Using this criteria E2 would then be the most evolved starburst. 
Observations of low quantum number RRLs (e.g. H43$\alpha$) tracing 
the dense gas are required to better constrain the recent star 
formation history.

The number of Lyman continuum photons required to 
ionize all four complexes, $\sim 16 \times 10^{52}$~s$^{-1}$, is 
a factor three lower than the value estimated by \citet{Ach95}.
The sum of the H92$\alpha$ line flux densities from the four modeled 
regions represents only $\sim$50\% of the total line flux density;
our analysis is thus restricted to the areas where the H53$\alpha$ line is detected. 
Considering this fact, the Lyman continuum rate obtained from 
the models is a lower limit for the total Lyman continuum rate emission in 
the center of M82.
Then, we conclude  that our results are in reasonable agreement 
with previous results.

M82 differs from Arp 220 in the sense that very 
high-density HII regions are not required to explain the
line and continuum emission from the center of M82. 
The SFR depends on the initial mass function and mainly on the 
upper limit used for the mass of the stars that are formed.
Using the Miller-Scalo IMF and mass limits of 1 and 100 M$_{\odot}$, 
the total SFR derived for Arp~220 
is $\sim 240$~M$_{\odot}$~yr$^{-1}$ \citep{An00}, whereas that for M82 
is $\sim 100$ times lower ($\sim 3$~M$_{\odot}$~yr$^{-1}$). 
Another estimate of the SFR for M82 using the same mass limits, the same IMF and assuming that
all the radio continuum emission is thermal gives an upper limit of 
$46$~M$_{\odot}$~yr$^{-1}$ \citep{Kr85}.
If the Salpeter IMF is used then the total SFR in M82 would be $\sim 3$ times larger.
The present mass of gas in the central region of M82 (160~pc) is 
estimated to be $\sim 2 \times 10^{8}$~M$_\sun$, a factor 30 lower
compared to Arp~220. 

\section{DISCUSSION}

\subsection{Radio Continuum and RRLs H92$\alpha$ and H53$\alpha$.}

There are two possible reasons that could explain why some 
continuum features were detected only at 8.3~GHz:
(i) the features are HII regions and the higher noise level 
at 43~GHz limits the detection,
(ii) the features are SNR with non-thermal spectrum 
($\alpha < 0$, $S_{\nu} \propto \nu ^{\alpha}$)
and flux density at 43~GHz $< 3 \sigma_{rms}$. For all the 
features identified as SNR, the 
spectral index value determined from our observations is 
in good agreement with a non-thermal spectra.
From 17 sources identified as HII regions, 13 sources have 
spectral index values in agreement with a thermal spectrum 
( $8.3- 43$~GHz).
The compact sources 42.69+58.2, 42.56+58.0 and 42.48+58.4 
are observed in a larger complex region; because this complex 
region is embedded in a more extended emission region with 
a similar flux density level,  we were not able to determine 
the continuum flux densities for each of them individually. 
The spectral index from this complex region is not in 
agreement with a thermal spectra; the contribution 
of synchrotron emission from the SNR at 8.3~GHz may explain this result.
Due to the lower angular resolution of our observations, 
the flux densities of some features are larger than previously 
reported measurements at 3.6~cm based on higher angular resolution data 
\citep{Hu94,All98}.
The larger measured flux densities arise from extended features 
that were resolved out by the higher angular resolution observations.
There are three compact sources that have been classified as 
SNRs and have detectable RRL emission.
The H92$\alpha$ line emission that seems to arise from SNRs 
may be accounted for if a group of HII regions is located  along 
the same line of sight as the SNR. 
The RRL emission that arise in the foreground ionized gas could even
be stimulated by the external emission arising from the SNR.
Quite possibly the externally stimulated emission plays an important 
role in these lines of sight.
For the source 41.95+57.5, the flux density at 8.3~GHz is 
13.6$\pm$0.4~mJy using observations made in 2001.
\citet{All98} at 8.4~GHz determined a value of 26.76$\pm$0.5~mJy 
in 1994-1995.
The implied decay rate of this radio supernova is in good agreement 
with the estimates (8.8\%~yr$^{-1}$) of \citet{All98}.

\subsection{Comparison with other tracers of ionized gas.}

As observed at an angular resolution of $2''$, the structure 
of M82 is very similar in the H92$\alpha$ line (Fig.~5) and 
the [Ne~II] line (Fig.~1 of Achtermann \& Lacy 1995).
In both cases the most prominent source is W1 (see Fig.~7), 
located $\sim 5''$ E of the 2.2~$\micron$ peak.
In the H92$\alpha$ line images (Fig.~5), a quasi-circular 
``hole'' with a FWHM deconvolved size of $\sim 2''$ is 
observed. The center of this hole is located at 
$\alpha=09^{h} 51^{m} 43\rlap.{^s}5$, $\delta=+69^{\circ} 55' 00\rlap.{''}4 $, 
which is close to the 2.2~$\micron$ peak determined by \citet{Di86} at
$\alpha=09^{h} 51^{m} 43\rlap.{^s}5$, $\delta=+69^{\circ} 55' 00\rlap.{''}7 $ 
and shifted from the one determined by \citet{Les90} at
$\alpha=09^{h} 51^{m} 43\rlap.{^s}6$, $\delta=+69^{\circ} 55' 00\rlap.{''}1 $ 
by $1\rlap.{^{''}}5$ in RA and by $0\rlap.{^{''}}3$ in Dec.
The position of this peak needs to be confirmed 
to definitively conclude if this hole as observed in H92$\alpha$ 
is, in fact, centered on this peak.
The regions E1 and W1 are connected by 
a faint ridge of emission on the N
side of this hole (see Figure~\ref{figlandc}).
Based on the [Ne~II] observations, \citet{Ach95} 
have suggested the presence of an 
ionized ring with a projected axial ratio of about three, a 
major axis of $11''$ at a position angle of $70^{\circ}$ and centered at 
$\alpha=09^{h} 51^{m} 43\rlap.{^{s}}4$, $\delta=+69^{\circ} 55' 00\rlap.{''}1$. 
Since the emission of H92$\alpha$ line is not affected 
by extinction, the lack of emission on the S side 
indicates that the ionized gas is not uniformly
distributed along this ring. 
In agreement with the conclusions of \citet{Ach95}, 
the observed brightness of E1 and W1 cannot be 
explained simply by a limb brightening effect.

At $0\rlap.{''}9$ resolution (Fig.~4) the spatial distribution 
of the H92$\alpha$ line is quite patchy. The region W1 is clearly 
decomposed into two main components: the group composed of W1b 
and W1c is the brightest and is located at the SE side and the 
group composed of W1e and W1f is located at the NW side. 
These two groups would be located on the S and the 
N side (respectively) of the major axis (PA~$70^{\circ}$) of the ring 
proposed by \citet{Ach95}. 
These two groups, as observed in the $Br\gamma$ ($2.17~\micron$) line, 
are separated by a ridge of extinction that extends along the major 
axis of M82 \citep{La94}.
Since the H92$\alpha$ line is not affected by extinction, 
the lack of emission between these two groups on the W side 
of the 2.2~$\micron$ peak is not due to dust extinction. 
On the E side, as observed in the H92$\alpha$ 
line, E1 appears relatively more diffuse than W1, with the 
same extension as the $Br\gamma$ line. 
This region is located on the S edge of a ridge of 
extinction that extends along the major axis of M82.
As we do not detect significant H92$\alpha$ line emission on 
this ridge, the distribution of $Br\gamma$ line emission for 
the region E1 does not seem to be significantly affected by 
dust extinction. 
The extinction derived from the Br$\gamma$ line was used 
to determine the orientation of M82. However, the 
distribution of the $Br\gamma$ line emission 
is not severely affected by dust extinction. Thus, 
the fact that the brightest $Br\gamma$ sources appear on the S
side cannot be used as a reliable indicator to determine the 
orientation of the galaxy \citep{La94}.

\subsection{Kinematics of the ionized gas.}

The kinematics in the central part of M82 have been 
extensively studied using observations of molecular 
lines, HI in absorption and the infrared lines 
$Br\gamma$ and [Ne~II]. 
From these studies a picture emerges in which there 
is an inner ring of ionized gas $\sim 11''$ ($\sim 200$~pc) 
in diameter possibly surrounded by a ring of molecular gas. 
An alternate model for the molecular distribution with two 
lobes has also been proposed by \citet{La94}.
The presence of a stellar bar, about one kiloparsec in length, 
is suggested by the $2.2~\micron$ light distribution \citep{Les90}. 
The ring of ionized gas could be related to x2 orbits at the
 center of the galaxy; observationally it is unclear if 
this ring is circular or has an oval distortion, since M82 is 
viewed highly inclined (i=$80^{\circ}$). \citet{Ach95} have shown that their 
data are consistent with a simple circular ring of about 150~pc 
diameter with a rotation velocity of $\sim 112$~km~s$^{-1}$ and 
possibly an associated outflow.
They also present a model involving x1 orbits with 
an almost end-on view of elliptical x2 orbits.

Along the major axis of M82 the $2.2~\micron$ light distribution 
is characterized by two plateaus, an inner plateau extending over 
$\sim 20''$ and an outer one extending over one arcmin; the outer 
plateau has been interpreted as a possible evidence for a stellar 
bar \citep{Te91}. 
Using images of M82 at the I, J, K and L' bands, 
\citet{La94} inferred that a dust lane is in front of the 
stellar population to the W of the nucleus, lying behind the 
stars to the E. This geometry is consistent with a stellar 
bar with leading dust lanes. 
\citet{Gre02}, based on extinction and polarization  
arguments, suggested that M82 is observed from below.


\subsubsection{The presence of a stellar bar.}

The total  H92$\alpha$ line distribution extends 
over $\sim 35''$ along the major axis of M82 (Fig ~5), covering the 
inner plateau and also the inner edges of the outer plateau. 
Since the resulting velocity field provides the
average velocity value of the gas at each position,
we have used the data cube to search for 
 the steepest velocity gradient in the center of M82.
The steepest velocity gradient is determined from
terminal velocities ($\sim 36$~km~s$^{-1}$~arcsec$^{-1}$) and
is found along a PA of $\sim 57^{\circ}$.
Only two regions, separated by $\sim 7\rlap.{''}8$, define this velocity gradient:
one of these regions is Ca and the other is 
a region located $0\rlap.{''}4$ from W1c.
The center defined by these two regions is located 
$1\rlap.{''}7$ E from the 2.2~$\micron$ peak determined by \citet{Di86}.
The velocity gradient could be alternatively defined between the region Ca and 
a region located near the center of E1a; in this case a velocity gradient of 
$\sim 27$~km~s$^{-1}$~arcsec$^{-1}$, at a PA of $\sim 68^{\circ}$ is obtained.
This PA agrees with the major axis of M82 determined on large scale. 
The position at mid-distance between these two regions 
(separated by $\sim 10''$) has an offset of
$\Delta_{\alpha}=-1\rlap.{''}5$, $\Delta_{\delta}=0\rlap.{''}4$ relative to 
the 2.2~$\micron$ peak \citep{Di86}. 
This offset corresponds approximately to the distance between the 2.2~$\micron$ 
peak and the center of the inner plateau observed in the $2.2~\micron$ profile 
(see Fig. 2 of Larkin et al. 1994). 
This asymmetry, with a truncation of the inner bar on 
the E side of the $2.2~\micron$ profile, has already 
been noticed by \citet{La94} and could reveal the presence of ionized gas in two lobes;
this suggestion is supported by our H92$\alpha$ observations since the regions Ca and  
E1a appear as the only significant features in the channels at 313 and 284~km~s$^{-1}$.
Although different velocity gradients are obtained using the regions
near E1a and W1c, both regions are located on the ridge of extinction 
determined by \citet{La94}. 
In any case, the center position defined by these two regions
is very close to the location of maximum extinction along this ridge.
Dust lanes are expected to be present at the intersection of x1 and x2 orbits.
This result suggests the presence of elongated x2 orbits.
Assuming that the PA of the x1 orbits corresponds to the outer plateau 
is $\sim 70^\circ$ \citep{Te91}, the major axis of the 
x2 orbits would be close to an end-on view.

The position-velocity ({\it p-v}) diagram,
using the integrated flux density 
along the minor axis direction by taking 
slices at a PA of 70$^{\circ}$, is shown in Figure~\ref{figpv} (top).
This PA was used in order to compare with results from [Ne~II] 
and Br$\gamma$ observations \citep{Ach95,La94}.
In the H92$\alpha$ line image there is emission at 
``forbidden'' velocities (near $-1\rlap.{''}8$ 
at 215~km~s$^{-1}$), inconsistent with purely circular motions.
If non-circular motions are present, additional velocity 
components are observed.
This feature is also observed in the $Br\gamma$ 
line images (Fig. 5 of Larkin et al.~1994). 
In [Ne~II], there is also emission at ``forbidden'' 
velocities but this is distributed in a 
symmetrical fashion about 200~km~s$^{-1}$ around the 2.2~$\micron$ peak.
The agreement of the H92$\alpha$ {\it p-v} image is closer 
to the $Br\gamma$ \citep{La94} {\it p-v} images than 
to the [Ne~II]  {\it p-v} image. 

Figure~\ref{figpv} (bottom) shows the {\it p-v} image constructed along the 
major axis using the PA of $65^{\circ}$.
In this case the profiles are in general relatively 
narrower when they are compared with those in Fig.~\ref{figpv} (top);
the binning along the minor axis is
over a smaller region of $\sim 1\rlap.{''}5$ as compared with $9''$. 
However, the spectra remain relatively broad except near $8''$ on the 
W side and $10''$ on the E side. 
These two regions are located at the transition zones between 
the inner and outer 2.2~$\micron$ plateaus. 
The asymmetry observed at $2''$ angular resolution, relative to 
the position of the $2.2~\micron$ peak, is still present for 
this PA=65$^{\circ}$, suggesting that the asymmetry is very likely
to arise in the bar itself.
In this case, there is no emission at ``forbidden'' 
velocities, indicating that the orbits must be confined 
to a relatively thin plane not observed edge-on. 
By comparing with optical measurements \citep{Mc93,Gre02},
we observe systematic effects:
in the inner part ($\leq  6''$) the terminal H92$\alpha$ 
velocities are close to the optical velocities only on the 
W side. Neither the gas nor the stars, as observed in the optical,
 have velocities near $\sim 325$~km~s$^{-1}$ from $+1''$ to $+6''$. 
On the other hand, there are no optical velocities near 
$\sim 200$~km~s$^{-1}$ from $-1''$ to $-6''$. Beyond $\pm 11''$, 
on the outer 2.2~$\micron$ plateau, the optical velocities 
are systematically closer to the low velocity component of 
the H92$\alpha$ profiles. These results would indicate 
that for the regions located in the outer plateau, the gas as 
well as the stellar emission observed in the optical comes 
preferentially from the leading edge of the bar in the x1 
orbits (see  Fig. 2 of Greve et al. 2002). 

The symmetry observed in the terminal velocities of the 
H92$\alpha$ line would indicate that 
the component of the ionized gas that we observe is not only 
in the x1 orbits but also in x2 orbits ($\le \pm 3''$) 
and/or with some ``spraying'' (at radial distances from 
$3''$ to $6''$). 
The result of gas shocking
in the intersection of x1 and x2 orbits is named by some authors 
as ``spray'' orbits (e.g. Downes et al.~1996).
The encounter of gas moving in these
two orbits breaks the flow along the x1 orbit into a spray that 
traverses interior of the bar until it reaches the far side of the bar. 
A schematic diagram for these spray orbits is shown by \citet{Do96}.
In this context it is difficult to explain why the optical 
velocities are close to the H92$\alpha$ terminal velocities 
only on the W side. One possibility would be that the gas
observed in the optical does not spray on the E side and the 
gas emitting the H92$\alpha$ line has orbits with radii which 
extend relatively closer to the nucleus. An alternate possibility 
would be that the x2 orbits are in a plane closer to $81^{\circ}$ 
inclination than the x1 orbits, the far side of the x2 orbits 
emitting in the optical being more severely affected by the extinction from dust 
in the plane of the x1 orbits. 

There is a thin transition edge between the
inner and outer plateaus not only observed in the {\it p-v} images 
made along the major axis (Fig.~\ref{figpv}) but also in the 
velocity dispersion image (Fig.~8b).
In Figure~8b an ellipse (fitted through 
the regions with local minima) is shown.
This ellipse is centered at the same 
position as the ring found by \citet{Ach95}, with a major 
axis of $17''$ at a PA of 72$^\circ$ and the same axial ratio of $\sim$3.5. 
This result agrees with the 2.2~$\micron$ light distribution 
for the inner plateau, consistent with a picture in which 
the inner and outer plateaus correspond to different families of orbits.     

\subsubsection{The inner ring.}

\begin{figure}[!ht]
\epsscale{0.8}
\plotone{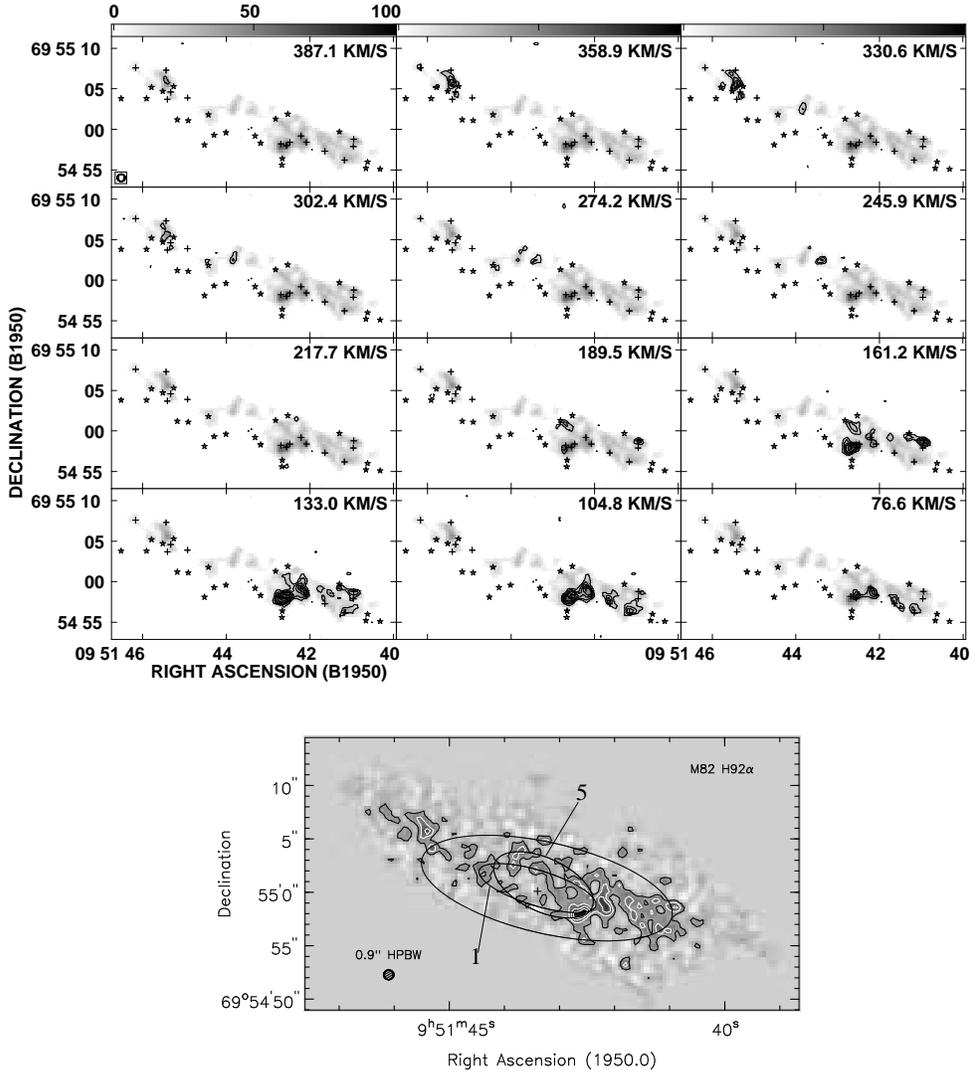}
\caption{Top) Contour images of the H92$\alpha$ line intensity throughout the velocity region
where emission was observed at intermediate-resolution ($0\rlap.{''}9$). 
The contour levels are $-$3, 3, 4, 5, 6, 7, 8, 9 times the rms noise of $\sim$0.07~mJy~beam$^{-1}$. 
The integrated line emission obtained from the low-resolution ($2''$) line images is shown in gray scale.
The gray scale image covers the range from 0 to 6~Jy~beam$^{-1}$~m~s$^{-1}$.
The marks indicate the position of the HII regions (crosses) and the SNRs (stars).
The central heliocentric velocity is given above each image. 
The beam size is shown in the bottom left corner of the first image.
Bottom)  Integrated H92$\alpha$ line emission at $0\rlap.{''}9$ angular resolution toward M82. 
Contour levels are 0.4, 0.8, 1.2, 1.6, 2.0~mJy~beam$^{-1}$~km~s$^{-1}$. 
The cross indicates the 2.2~$\mu$m peak \citep{Les90}.
The outer ellipse shows the extension of the inner 2.2 $\micron$ plateau.
The ellipses shown correspond to models of Table 6 in row 1 and 5, respectively. 
}
\label{figh92ir}
\end{figure}

\begin{figure}[!ht]
\plotone{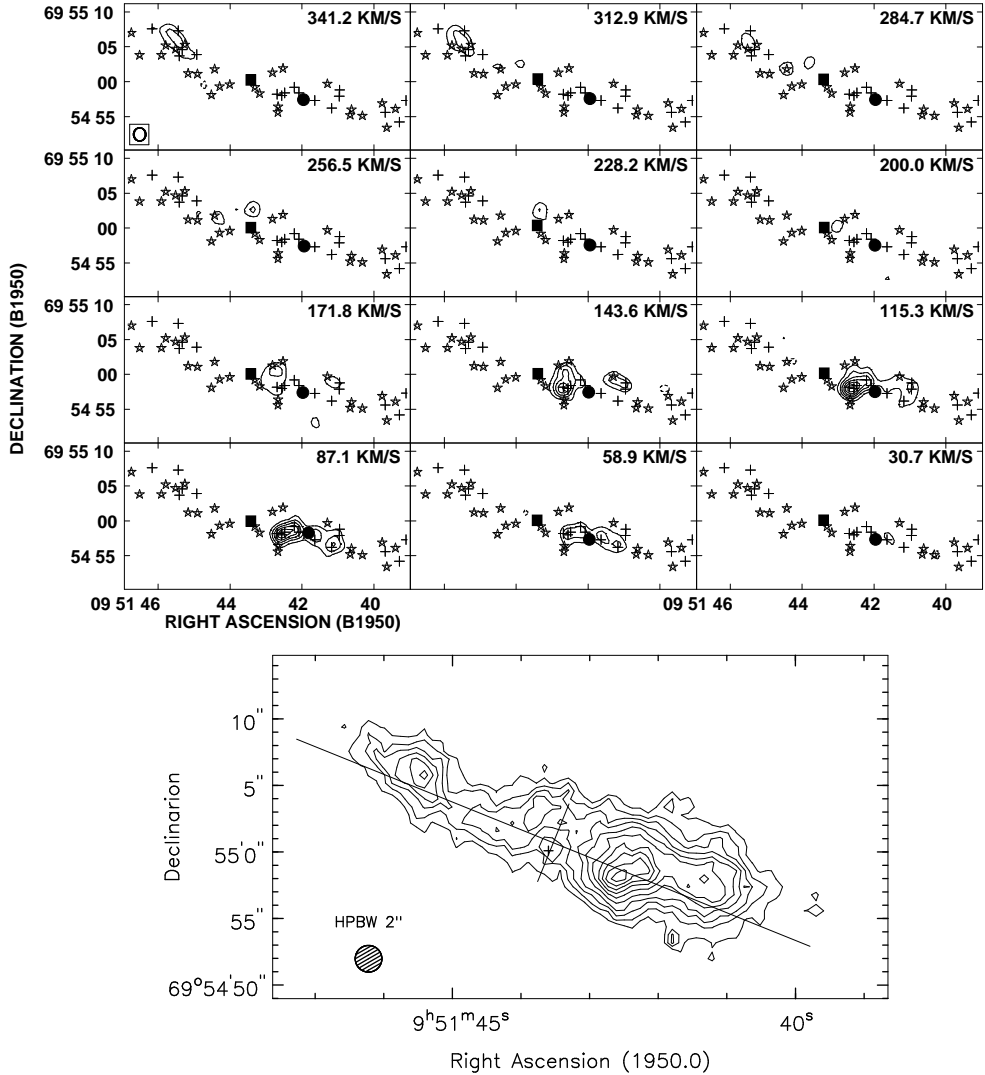}
\caption{Top) Contour images of the H92$\alpha$ line at low angular resolution ($2''$). 
The contour levels are $-$3, 3, 5, 7, 9, 11, 13, 15, 17, 19 times the rms noise of
$\sim$0.07~mJy~beam$^{-1}$. 
The filled circle shows the position of G41.95+57.5 and the 
filled square the peak at 2.2~$\mu$m \citep{Les90}.
The beam size is shown in the bottom left corner of the first image.
Bottom) Integrated H92$\alpha$ line emission at $2''$ angular resolution toward
M82. Contour levels are from 15.82 up to 158.2 by increment of
15.82~mJy~beam$^{-1}$~km~s$^{-1}$.  The two line segments
show the minor and major axis of a thin disk which would be observed with an
inclination of 81$^\circ$ and centered at the position of the 2.2~$\mu$m
peak \citep{Di86}. The cross indicates the 2.2~$\mu$m peak \citep{Les90}.}
\label{figh92lr}
\end{figure}

\begin{figure}[!ht]
\epsscale{0.5}
\plotone{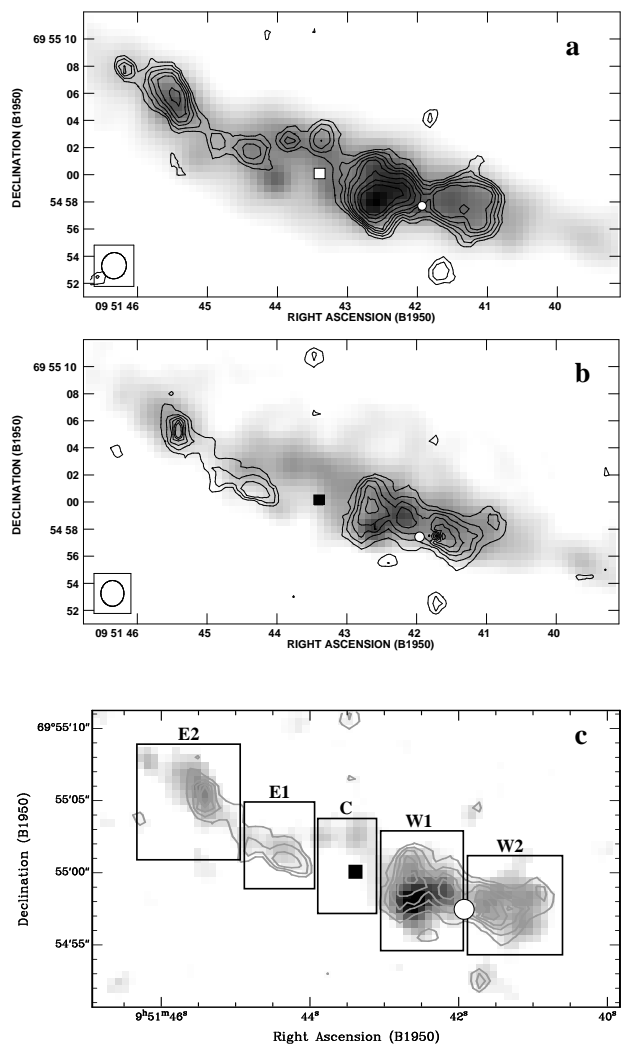}
\caption{
a) The integrated H92$\alpha$ line emission (contours) superposed 
on the radio continuum at 8.3~GHz (gray-scale) at $2''$ angular resolution.
Contour levels are drawn at 18 through 148~mJy~beam$^{-1}$~km~s$^{-1}$ in steps of
26~mJy~beam$^{-1}$~km~s$^{-1}$.
The gray scale for the continuum covers the range from 5.4 to 55~mJy~beam$^{-1}$.
b) The integrated H53$\alpha$ line emission (contours) superposed 
on the radio continuum image of M82 at 43~GHz (gray-scale) at
$2''$ angular resolution. 
Contour levels are at 57, 114, 171, 228, 285~Jy~beam$^{-1}$~km~s$^{-1}$.
The gray scale for the continuum covers the range from 2.3 to 34~mJy~beam$^{-1}$.
c) Integrated intensity of H92$\alpha$ line (gray-scale) and
integrated intensity image of the H53$\alpha$ line  (contours).
Contour levels are 57, 114, 171, 228~Jy~beam$^{-1}~$~km~s$^{-1}$.
The gray scale covers the range from 8.4 to 196~mJy~beam$^{-1}$~km~s$^{-1}$.
The square marks the position of the 2.2~$\micron$ peak \citep{Les90}
and the circle marks the position of the SNR G41.95+57.5.}
\label{figlandc}
\end{figure}

\begin{figure}[!ht]
\epsscale{0.80}
\plotone{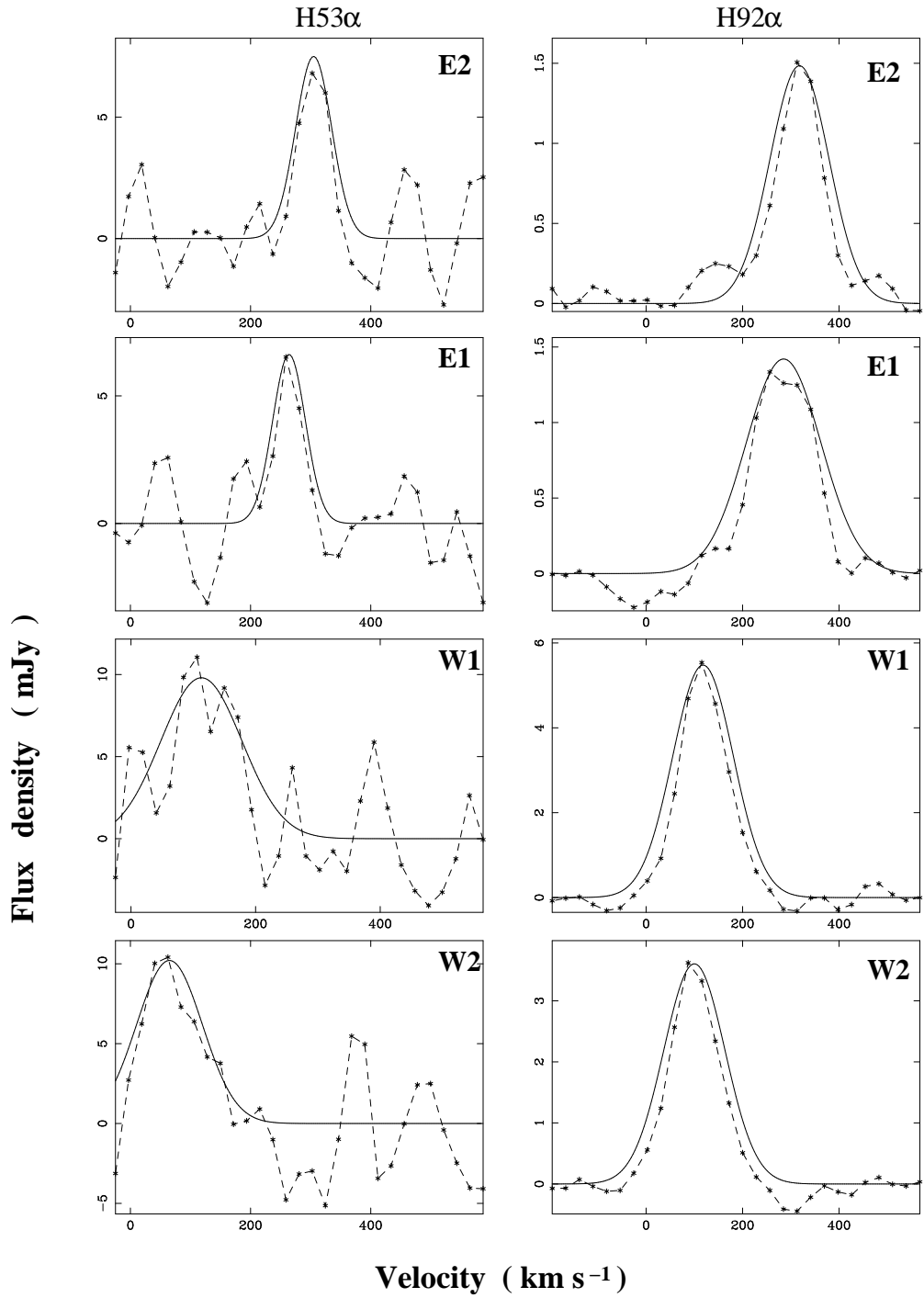}
\caption{H92$\alpha$ and H53$\alpha$ spectra toward the regions labeled in Figure~\ref{figlandc}c.
The dashed curves are the observed intensities and the solid lines
represent the fitted gaussians as a function of heliocentric velocity.}
\label{figspec}
\end{figure}

\begin{figure}[!ht]
\epsscale{1.0}
\plotone{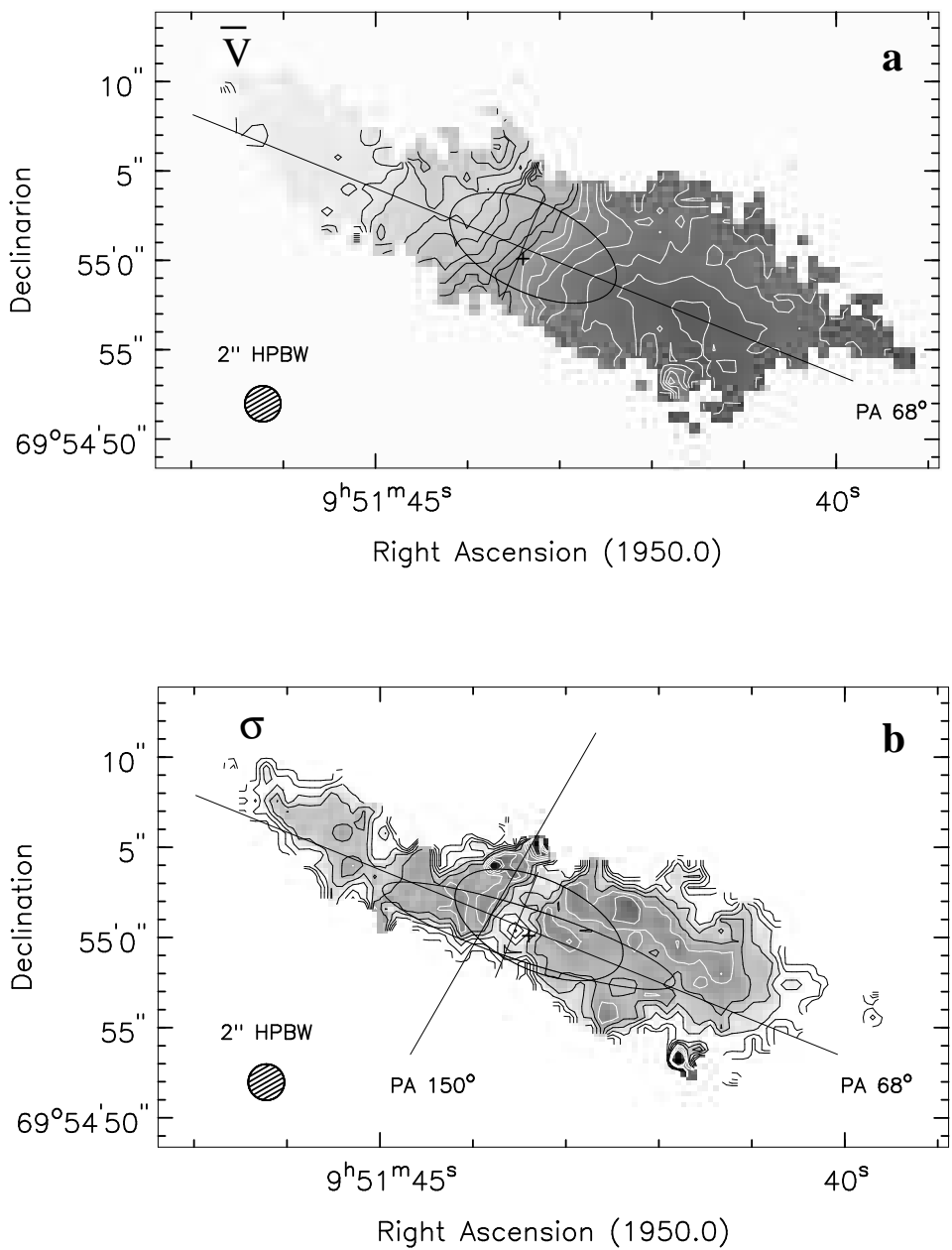} 
\caption{
a) H92$\alpha$ velocity field at $2''$ angular resolution. Contour intervals
are 16~km~s$^{-1}$, the iso-velocity contours $\ge$ 200 km~s$^{-1}$ are in black. 
b) Spatial distribution for the velocity dispersion of the H92$\alpha$
profiles. The contours from 15 to 35~km~s$^{-1}$ 
are in black and above this level in white. 
The contour interval is 5~km~s$^{-1}$. The most elongated 
ellipse is a fit to a ridge with minimum velocity 
dispersion which corresponds to the transition 
between the inner and outer plateau in the 
2.2~$\micron$ brightness distribution. 
In both images, the less elongated ellipse corresponds to the model of row 5 (Table 6).
The two perpendicular line segments are the same as in 
Figure~\ref{figh92lr} (bottom). The cross indicates the center of the ring fitted 
by \citet{Ach95}.}

\label{figvf}
\end{figure}

\begin{figure}[!ht]
\epsscale{0.8}
\plotone{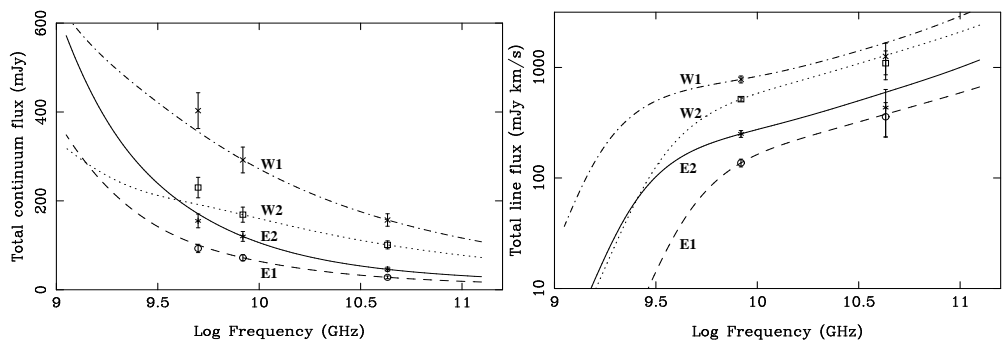}
\caption{
Expected variation of radio continuum (left) and integrated line
flux density (right) for four of the regions identified in Fig.~\ref{figlandc}c.
The continuum flux density has been computed using the model of 
multiple HII regions with a single component
plus non-thermal emission arising from SNR. The integrated 
line emission curves correspond to the single-density models of Table 5.}
\label{figmods}
\end{figure}

\begin{figure}[!ht]
\epsscale{0.80}
\plotone{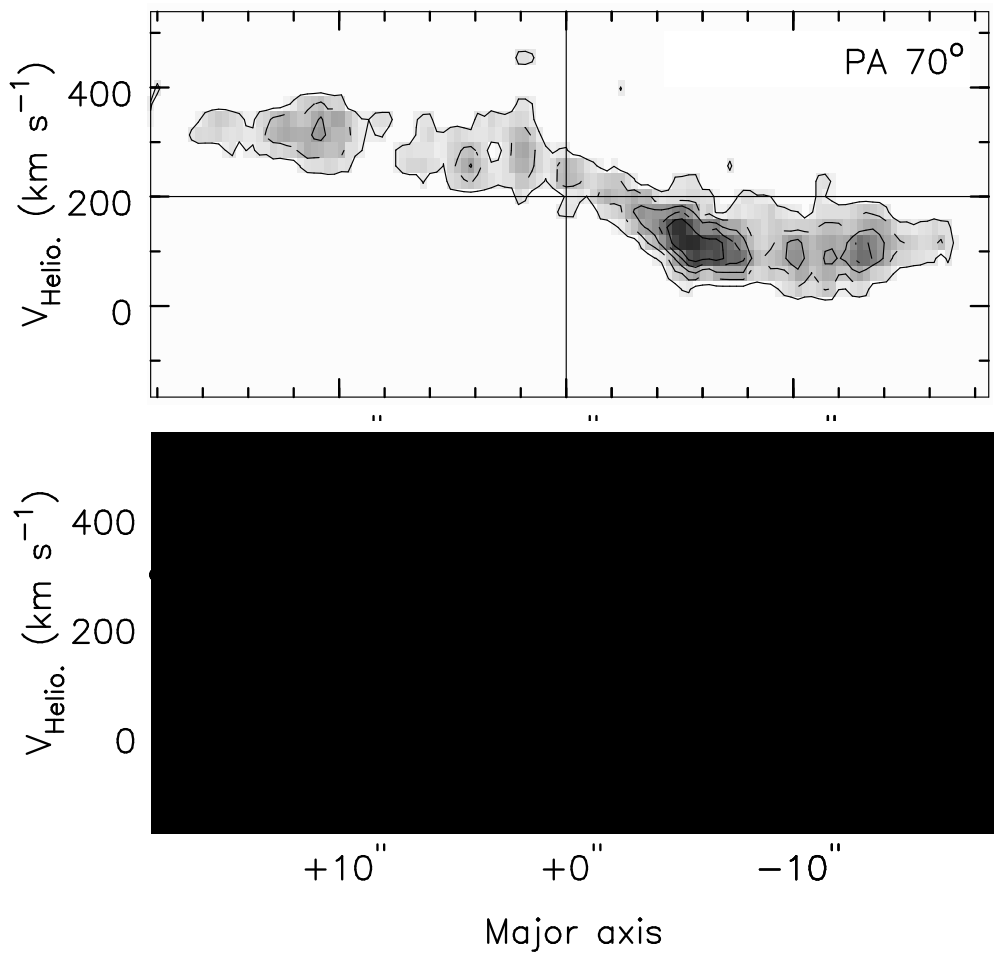} \caption{ 
H92$\alpha$ position-velocity diagram summed over the minor axis 
(angular resolution $0\rlap.{''}9$), along:
top) the main axis of the stellar bar with PA$=$70$^{\circ}$. 
Contours are from 0.3 to 3.0 in steps of 0.6~mJy~beam$^{-1}$, and
bottom) the major axis of M82 determined on larger scales with 
PA$=$65$^{\circ}$.
Small circles marks the velocity of the stars as determined from 
[Ca~II] line \citep{Mc93} and big circles show the velocity of the 
gas determined in Pa(10) line \citep{Mc93}. Contours are
0.1, 0.4 and 1.5~mJy~beam$^{-1}$. The beam is $2\rlap.{''}5 \times 0\rlap.{''}9$.
The gray scales follow the contours.
The slices were centered at $\alpha$=09$^{h}$51$^{m}$43$\rlap.{^{s}}6$, 
$\delta$=69$^{\circ} 55'00''$.}
\label{figpv}
\end{figure}

A description in terms of elongated x1 and x2 orbits is not 
the only model that can explain the observed kinematics in M82. 
The observed velocities distribution is not consistent
with a rotating disk unless it consists of a set of rings 
with different tilt angles which may be contracting.
Different methods to fit a ring in the center of M82 were used;
the results are listed in Table~6. 
Row~1 of Table~6 lists the results of \citet{Ach95}, 
column~1 lists the central position of the ring, column~2 the 
position angle of the major axis, columns~3 and 4 the angular size of 
the major and minor axis, column~5 the systemic velocity, column~6 
 the rotation velocity, column~7 the expansion velocity,
column~8 the dispersion about the model and column~9 
the spectral line used to obtain the parameters of the ring.
\citet{Ach95} fixed the geometry of the ring based on the 
physical appearance of the distribution of the [Ne~II] line 
emission and then fitted the velocity pattern along that ring,
 assuming that the ring lies in the plane of the galaxy,
 centered on the kinematical center.
Row~2 is the result obtained from the H92$\alpha$ 
velocity pattern (Fig.~\ref{figvf}a), using the same 
morphological parameters listed in row 1. 
Following the same procedure for the H92$\alpha$ line emission, 
the ellipse fitted by \citet{Ach95} would need to be shifted 
$\sim 0\rlap.{''}3$ to the N
(Fig.~\ref{figh92ir}) to coincide with the ridge of emission.
In row~3, this shift of $0\rlap.{''}3$ has been applied.
In row 4, the position of the center is a free parameter.
Finally, row 5 lists the parameters of the ring 
(shown in Fig.~\ref{figh92ir}) obtained
without imposing any morphological constraints;
this procedure gives the solution with the lowest residual dispersion.

Based on the spatial distribution of the line emission,
the existence of the ring is not obvious in any of the models presented in Table~6.
According to the ring model of \citet{Ach95} and the one
described by row 5 (Table~6), the rotation velocity $V_{rot}$ has a first maximum 
at a radius of $\sim 5''-6''$.
In the model of \citet{Ach95} the ring surrounds a region 
 devoid of ionized gas, in contrast with the 
model presented here (the best fit to the velocity pattern). 
Both models require not only a contraction velocity but also 
that the inclination angle of the rings is different than the 
inclination of the plane of the outer part of M82 (i=$81^\circ$).
The major axis of these rings is very close to the PA determined 
from the large scale morphology (68$^\circ$). 

\begin{figure}[!ht]
\epsscale{1.0}
\plotone{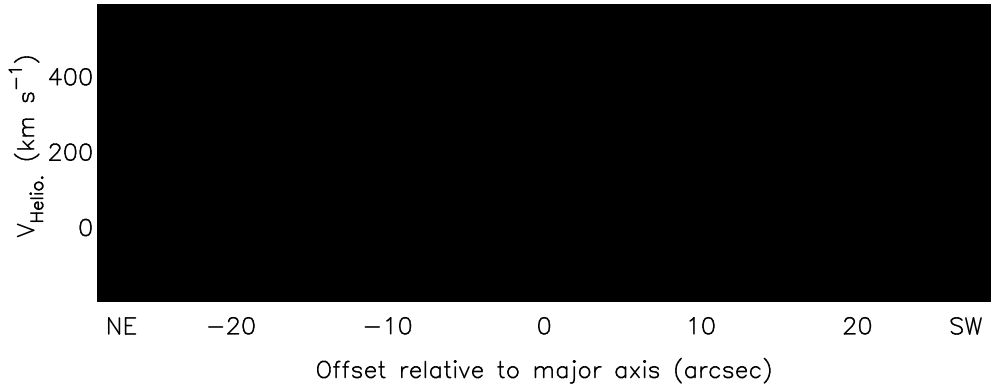} 
\caption{Position-velocity image along the E ridge of 
broad profiles (see Fig.~\ref{figvf}) at 150$^{\circ}$. The angular
resolution is $0\rlap.{''}9$. 
The offsets are relative to the major axis,
NW to the left and SE to the right, the origin being at
$\alpha=$09$^h$51$^m$43$\rlap.{^{s}}$81, $\delta=$69$^{\circ}$55$'$1$\rlap.{^{``}}$1.
Contours are -0.1, -0.05~mJy~beam$^{-1}$ (dashed) and from 0.05 to 0.5~mJy~beam$^{-1}$
 in steps of 0.05~mJy~beam$^{-1}$.}
\label{figof}
\end{figure}

In the velocity dispersion image (Fig.~\ref{figvf}b), 
a well defined band at a PA of $\sim 150^\circ$ is observed with
systematically narrower line profiles.
The length of this band is not restricted to the extent of
 the minor axis of the ring with axial ratio of $\sim 3.5$ 
\citep{Ach95}; in addition this band covers the extent of the 
velocity pattern along that direction. 
On both sides of this band, there is a ridge in which the broadest 
profiles are observed.
These regions of large velocity dispersions 
are explained by the terminal velocities V$_{t}$ with 
$\vert $V$_t -  V_{sys} \vert \simeq 120$~km~s$^{-1}$, along with a second 
velocity component in the spectra with velocities closer to V$_{sys}$, 
$\vert $V$_t -  V_{sys} \vert \simeq 50$~km~s$^{-1}$.
Figure~\ref{figof} shows the {\it p-v} diagram along the 
E ridge. The regions where V$_{t}$ $\ge +300$~km~s$^{-1}$ 
are not restricted to the range 
$\le \pm 1\rlap.{''}7$, which corresponds to the regions associated 
with the ring of axial ratio of $\sim$3.5 (see Table 6).
Outside the range $-5''$ to $+3''$, 
{\it i.e.} beyond the ring of axial ratio 2 (see Table 6), the bright 
component remains confined near $\sim 250$~km~s$^{-1}$.
We conclude that, if gas in x2 orbits and/or a fast rotating 
nuclear ring (spray between the x1 and x2 orbits) is 
responsible for V$_{t} > +300$~km~s$^{-1}$,
it cannot be confined in a thin plane which would give this 
axial ratio of $\sim 3. 5$.

\subsubsection{Outflow in the halo.}
Fig.~\ref{figof} shows that for offsets larger 
than $-5''$ there is emission at two different velocities,
one component emerging near $\sim150$~km~s$^{-1}$ and the second one
near $\sim 350$~km~s$^{-1}$. At $\sim -10''$ the emission is 
detected in the range $\sim250$~km~s$^{-1}$ and $\sim30 - 100$~km~s$^{-1}$. 
These results agree with the optical measurements 
\citep{Mc93} in the lines of [SII], [NII], H$\alpha$.  
A more detailed view of the radial velocities of the 
outflow, observed in H$\alpha$, are shown in Figure~10 
of \citet{Sh98}.
The H92$\alpha$ observations reveal the base of the outflow into
the halo. 
On the N side of M82 there are two velocity components: 
the ionized gas on the far side of the cone (component I) 
has radial velocities $\sim 250$~km~s$^{-1}$ 
and on the near side of the cone (component II) the gas has 
radial velocities in the range $\sim 30$ to $100$~km~s$^{-1}$. 
On the S side ($\sim +5''$), the velocities are confined to 
$\sim 200$~km~s$^{-1}$ (component IV), again in agreement with 
the optical measurements \citep{Sh98}.
The geometry of the outflow and description of these four components 
are shown in Figure 5 of \citet{Mc95}.

Due to the high inclination of M82, it is difficult to 
distinguish between an {\it ad-hoc} model of circular motions 
with radial contraction {\it vs} elliptical x1/x2 orbits.
The ring models indicate
contraction instead of expansion velocities (Table 6). 
The presence of different velocity components in the H92$\alpha$
line emission along a PA of $150^{\circ}$ indicates that
the ionized gas near the nucleus has already been perturbed by the outflow.
A barred potential seems to provide the mechanism to bring gas from 
the outer parts of the disk of M82 into the central part.
In this way, the gas flowing into the center of M82 would be
consumed in the starburst with some gas outflow into the halo. 

\subsection{Existence of an AGN}
The presence of an AGN in the center of M82 has been 
proposed by \citet{Ma99} based on X-ray data.
These authors suggest that there is an under-luminous 
AGN in the center of M82, comparable to Sgr A$^{*}$.
In the continuum at 8.3~GHz (Fig.~2a), we observe a 
source located near the 2.2~$\mu$m peak \citep{Les90}.
In the continuum at 43~GHz, we do not detect this 
source above the 3$\sigma_{rms}$ level.
\citet{Mu94}, using MERLIN observations at 5~GHz with 
angular resolution of 50~mas, report a compact source 
(43.55+60.0) that was identified as a young supernova 
remnant with shell-like structure.
Thus, no clear evidence has been found for the existence of 
a compact radio source that can be associated with
an AGN in the center of M82 within our 8.3~GHz $3\sigma_{rms}$ 
limit of 0.12~mJy~beam$^{-1}$.

\section{CONCLUSIONS.}

We have presented high angular resolution ($<1''$) VLA observations 
of radio continuum emission from M82 at 8.3 and 43~GHz
and of the H92$\alpha$ and H53$\alpha$ RRLs.
In the radio continuum images, we identified 19 new compact
sources at 8.3~GHz and five at 43~GHz.
Of these newly identified sources, four are SNRs, 
four are HII regions and the nature of the remaining 
13 is uncertain. Three SNRs show the presence of 
H92$\alpha$ RRL emission which can be interpreted as 
line emission produced by HII regions along the 
line of sight. 
No compact source is observed near the position of 
the derived kinematic center.
Thus, in agreement with \citet{Mu94}, we find that 
there is no compelling evidence for the existence of an AGN at 
the center of M82.

We have modeled the line and continuum emission using
a collection of HII regions.
We have considered models with a single density as well
as models with two density components.
The H53$\alpha$ RRL mainly traces the high-density ionized gas.
Models with multiple HII regions were used to reproduce the observations in 
the continuum at 5, 8.3 and 43~GHz as well as the H92$\alpha$ 
and H53$\alpha$ RRL emission from four separate HII complexes.
The two-density model is considered to be more appropriate
to model the line and continuum emission since compact continuum
sources have been identified already as HII regions.
However, as the two-density model is based on previous 
identification of thermal compact sources, the emission
from region E1 (no compact thermal sources 
have been identified) has been modeled using only the single-density 
model.
In general the low-density component is characterized by 
HII regions with sizes of $\sim$0.8~pc and an average 
electron density of $\sim 5 \times 10^{3}$~cm$^{-3}$ 
and the high-density component is characterized by HII 
regions with typical sizes of $\sim 0.15$~pc and an average 
electron density of $\sim 3.5 \times 10^{4}$~cm$^{-3}$.
From the two-density model, we can infer that the 
starburst in the region W1 is the youngest.
The derived mass of ionized gas in the central region of M82 is 
$\sim 2 \times 10^{8}$~M$_{\odot}$, 30 times lower than 
for Arp~220 \citep{An00}.
Based on the models, it has been inferred that
M82 has a star formation rate of $\sim 3$~M$_{\odot}$~year$^{-1}$,
a factor $\sim$ 100 lower compared with the merging system Arp 220.
For a more accurate determination of the properties of 
the ionized gas over a complete density range, high-resolution observations 
of RRLs at additional higher and lower frequencies are required.

The H92$\alpha$ line emission extends over $\sim 35''$. The
steep velocity gradient, measured at PA of 68$^{\circ}$, is 
26~km~s$^{-1}$~arcsec$^{-1}$ in the inner $\sim$100~pc of M82.
At a distance of $\sim$80~pc from the kinematic center, 
the gas has a maximum radial velocity of $\sim$150~km~s$^{-1}$, 
implying an enclosed total dynamical mass of $\sim$10$^{8}$~M$_{\odot}$.
From the {\it p-v} diagram of the H92$\alpha$ line, we 
observe deviations from circular motions.
The observed velocity pattern cannot be due to 
pure circular motions; apparently the orbits in the inner 
parts of M82 are more face-on than in the outer parts.
In the velocity dispersion image, the H92$\alpha$ line 
traces the base of a large scale outflow into the halo. 
This outflow is observed only on the N side.
We find the PA of the axis of the outflow to be 150$^{\circ}$. 
Along this axis, the widths of the profiles are narrow, suggesting 
collimated flow with small opening angle in the regions within the disk.
Within the disk, the steepest gradient ($\sim 36$~km~s$^{-1}$~arcsec$^{-1}$), 
is measured from the terminal velocities. This gradient is 
 observed at a PA of $\sim 57^\circ$, in a direction perpendicular to
the axis of the outflow into the halo.
This direction is not parallel to the PA of the large scale 
major axis of M82 ($\sim 65^{\circ}$) nor the PA of the 
2.2~$\micron$ bar ($\sim 70^{\circ}$).

The best fit with an {\it ad hoc} model of 
circular orbits leads to a contracting ring that has its
kinematic center shifted $\sim 1''$ W of the 
2.2~$\micron$ peak and located approximately 
in the middle of the 2.2~$\micron$ inner plateau.
The nodes characterizing this ring coincide 
with the dust lanes; this coincidence could represent 
additional evidence of gas in x2 orbits of a bar potential. 
Due to the presence of the outflow, the 
kinematics observed in the H92$\alpha$ line provide 
possible evidence for the presence of x2 orbits originating 
from a bar potential.
The inner and outer 2.2~$\micron$ plateaus correspond to two 
different families of orbits; the transition zone between the
two plateaus has a signature in the kinematics in the form of a reverse 
in the velocity gradient that has been observed near the major axis.
 
The National Radio Astronomy Observatory is a facility of 
the National Science Foundation operated under cooperative agreement 
by Associated Universities, Inc. We thank J. M. Torrelles for making 
available his H92$\alpha$ data. We also thank Yolanda G\'omez and 
anonymous referee for their very  
useful comments. CR acknowledges the support from UNAM and CONACyT, M\'exico.

\clearpage

\begin{deluxetable}{ccc}
\tablecolumns{3}
\tablewidth{0pc}
\tablecaption{P{\footnotesize ARAMETERS OF THE}  VLA O{\footnotesize BSERVATIONS OF} M82 }
\tablehead{
\colhead{Parameter} & \colhead{H92$\alpha$ Line} & \colhead{H53$\alpha$ Line} \\}
\startdata
Right ascension (B1950)\dotfill                 &       09$^{h}$51$^{m}$42$\rlap.{^{s}}$38 &   09$^{h}$51$^{m}$42$\rlap.{^{s}}$51\\
Declination (B1950)\dotfill                     &       69$^{\circ}$54$'$59$\rlap.{''}$2    & 69$^{\circ}$55$'$00$\rlap.{''}$0\\
Total observing duration (hr)\dotfill           &       32              &       13\\
Bandwidth (MHz)\dotfill                         &       25              &       50$\times$2\\
Number of spectral channels\dotfill             &       31              &       15$\times$2\\
Center V$_{Hel}$ (km~s$^{-1}$)\dotfill          &       200             &       160\\
Velocity coverage (km~s$^{-1}$)\dotfill         &       700             &       350$\times$2\\
Velocity resolution (km~s$^{-1}$)\dotfill       &       56              &       44\\
Amplitude calibrator\dotfill                    &       3C286           &       3C286\\
Phase calibrator\dotfill                        &       1044+719        &       0954+658 \\
Bandpass calibrator\dotfill                     &       3C48            &       1226+023\\
RMS line noise per channel (mJy/beam\tablenotemark{a} )\dotfill   &       0.07            &       1.3-2.0 \\
RMS, continuum (mJy/beam\tablenotemark{a} )\dotfill               &       0.04            &       0.5 \\
\enddata
\tablenotetext{a}{Synthesized beam size of $2''$.}
\end{deluxetable}

\hoffset -0.5in
\begin{deluxetable}{cccccccccccc}
\tabletypesize{\scriptsize}
\tablecolumns{11}
\textwidth 24cm
\tablewidth{0pc}
\tablecaption{
C{\footnotesize ONTINUUM FLUX DENSITIES, COMPACT FEATURES.}}
\tablehead{
 \colhead{}
& \multicolumn{3}{c}{8.3~GHz}& & \multicolumn{3}{c}{43~GHz} \\
 \cline{2-4} \cline{6-8} \\
\colhead{Source ID$^{1}$}       & \colhead{Coordinates$^{1}$}   & \colhead{Size$^{2}$}        & \colhead{S$_{8.3}$}   
                &       & \colhead{Coordinates$^{1}$}   & \colhead{Size$^{2}$}        & \colhead{S$_{43}$}    
                        & \multicolumn{2}{c}{Spectral index }    & \colhead{RRL} & \colhead{Type} \\

                        &                               & \colhead{($''$)}              & \colhead{(mJy)}       
                &       &                               & \colhead{($''$)}               & \colhead{(mJy)}              
                        & \colhead{8.3-43~GHz}  & \colhead{Low freq.}   & \colhead{feature}              \\
 \colhead{(1)}  & \colhead{(2)} & \colhead{(3)} & \colhead{(4)} && \colhead{(5)}        & \colhead{(6)} & \colhead{(7)} &
 \colhead{(8)}  & \colhead{(9)} & \colhead{(10)} & \colhead{(11)}       \\
}
\startdata
37.54+53.2 & 37.54+53.2$^{a}$  & 0.5$^{g}$   & 0.43$\pm$0.03 && $-$    & $-$   & $<$1.5        & $<$0.74
& $-$                           & $-$   &       \\
38.76+53.4  & 38.75+53.5$^{b}$  & 0.5$^{g}$   & 0.24$\pm$0.07 && $-$    & $-$   & $<$1.5        & $<$1.1
& 0.54$\pm$0.27$^{d}$           & $-$   & HII$^{f}$     \\
39.10+57.3 & 39.11+57.3$^{b}$   & u   & 3.81$\pm$0.04 && $-$    & $-$   & $<$1.5        & $< -$0.55
& $-$0.53$\pm$0.09$^{d}$        & $-$    & SNR$^{f}$     \\
39.29+54.2 & 39.27+54.1$^{b}$   & 0.4   & 3.6$\pm$0.1   && 39.25+54.1$^{b}$   & 0.5   & 3.5$\pm$0.3   & $-$0.02$\pm$0.02
& 1.64$\pm$0.16$^{d}$           & $-$    & HII$^{f}$     \\
39.40+56.1 & 39.43+55.9$^{b}$   & 0.5   & 1.9$\pm$0.1   && $-$          & $-$   & $<$1.5        & $< -$0.14
& $-$1.04$\pm$0.22$^{d}$        & $-$   & SNR$^{f}$     \\
39.64+53.4 & 39.65+53.2$^{b}$   & u   & 1.0$\pm$0.05  && $-$          & $-$   & $<$1.5        & $<$0.24
& $-$0.20$\pm$0.05$^{c}$        & $-$   & SNR$^{f}$     \\
39.68+55.6  & 39.67+55.3$^{b}$  & 0.4   & 3.9$\pm$0.2   && 39.66+55.4$^{b}$   & 0.5   & 3.6$\pm$0.5   & $-$0.05$\pm$0.09
& 1.03$\pm$0.09$^{d}$           & $-$   & HII$^{f}$     \\
39.77+56.9 & 39.78+56.9$^{b}$   & 0.4   & 0.7$\pm$0.05  && $-$          & $-$   & $<$1.5        & $<$0.45
& $-$0.50$\pm$0.06$^{c}$        & $-$   & SNR$^{f}$    \\
39.92+55.9 & 39.92+55.9$^{a}$   & 0.7   & 2.7$\pm$0.1   && $-$          & $-$   & $<$1.5        & $< -$0.35 
&    $-$                        & $-$   & SNR?$^{e}$       \\
40.10+55.0 & 40.15+54.5$^{b}$   & 0.7   & 0.6$\pm$0.05  && $-$          & $-$   & $<$1.5        & $<$0.54
& $-$                           & $-$   &       \\
40.32+55.1 & 40.33+55.1$^{b}$   & 0.4   & 0.8$\pm$0.04  && $-$          & $-$   & $<$1.5        & $<$0.37
& $-$0.23$\pm$0.21$^{d}$        & $-$   & SNR$^{f}$     \\
40.49+57.4 & 40.49+57.4$^{a}$   & 1.0   & 2.7$\pm$0.1   && $-$          & $-$   & $<$1.5        & $< -$0.35 
& $-$                           & $-$   & SNR?$^{e}$       \\
40.62+56.0 & 40.63+56.0$^{b}$   & 0.5   & 1.7$\pm$0.1   && $-$          & $-$   & $<$1.5        & $< -$0.01
& $-$0.72$\pm$0.25$^{d}$        & $-$   & SNR$^{f}$     \\
40.66+55.2 & 40.67+55.1$^{b}$   & 0.4   & 5.6$\pm$0.1   && 40.65+55.1$^{b}$   & 0.5   & 2.7$\pm$0.1   & $-$0.43$\pm$0.05      
& $-$0.54$\pm$0.08$^{d}$        & $-$   & SNR$^{f}$     \\
40.95+58.8 & 40.93+58.7$^{b}$   & 0.5   & 4.5$\pm$0.4   && 40.92+58.7$^{b}$   & 0.5   & 3.9$\pm$0.2   & $-$0.09$\pm$0.06
& 0.44$\pm$0.13$^{d}$           & W2h  & HII$^{f}$     \\
40.96+57.9 & 40.96+57.8$^{b}$   & u   & 2.3$\pm$0.2   && 40.96+57.6$^{b}$   & 0.5   & 1.8$\pm$0.1   & $-$0.15$\pm$0.05
& $>$1.2$^{+}$                  & $-$   & HII$^{f}$     \\
41.17+56.2 & 41.17+56.1$^{b}$   & 0.4   & 5.4$\pm$0.2   && 41.16+56.2$^{b}$   & 0.7   & 4.9$\pm$0.4   & $-$0.06$\pm$0.03
& 0.87$\pm$0.14$^{d}$           & W2f    & HII$^{f}$     \\
41.29+59.7 & 41.33+59.2$^{b}$   & 0.7   & 3.0$\pm$0.2   && $-$          & $-$   & $<$1.5        & $< -$0.41 
& $-$0.47$\pm$0.09$^{d}$        & W2e    & SNR$^{f}$     \\
41.62+59.9 & 41.62+59.9$^{a}$   & 0.5   & 1.6$\pm$0.1   && $-$          & $-$   & $<$1.5        & $< -$0.04 
& $-$        & $-$   &        \\
41.95+57.5 & 41.95+57.4$^{b}$   & u   & 13.6$\pm$0.4  && 41.95+57.5$^{b}$   & 0.7   & 5.0$\pm$0.1   & $-$0.60$\pm$0.02
& $-$0.80$\pm$0.05$^{c}$        & $-$   & SNR?$^{f}$    \\
42.08+58.4 & 42.11+58.3$^{b}$   & 0.4   & 4.0$\pm$0.2   && 42.09+58.3$^{b}$   & 0.5   & 3.6$\pm$0.2   & $-$0.06$\pm$0.04 
& 1.32$\pm$0.17$^{d}$           & $-$   & HII$^{f}$     \\
42.21+59.0 & 42.20+59.0$^{b}$   & 0.5   & 6.7$\pm$0.1   && 42.21+59.0$^{b}$   & 0.7   & 6.5$\pm$0.3   & $-$0.02$\pm$0.03
& 1.16$\pm$0.13$^{d}$           & W1f    & HII$^{f}$     \\
42.67+55.6 & 42.67+55.5$^{b}$   & u   & 0.92$\pm$0.04 && $-$          & $-$   & $<$1.5        & $<$0.29
& $-$1.3$\pm$0.2$^{d}$          & $-$   & SNR$^{f}$     \\
42.69+58.2 $\lceil$  &          &       &       &&              &       &               & 
&  $\rceil$ 1.04$\pm$0.13                       &  W1c   & HII$^{f}$     \\ 
42.56+58.0 $\vert$   & 42.64+57.9$^{b}$       & $\sim 1.4$ & 32.4$\pm$0.7     && 42.64+57.9$^{b}$   & $\sim 1.2$    & 16.2$\pm$0.8  
& $-$0.41$\pm$0.2 &  $\vert$ 0.88$\pm$0.14      &  W1c   & HII$^{f}$   \\
42.48+58.4 $\lfloor$ &          &       &       &&              &       &               &  
& $\rfloor$~~~~$>$1.2~~~~                       &  W1c   & HII$^{f}$     \\
43.00+59.0  & 43.02+58.9$^{b}$  & 0.4   & 3.9$\pm$0.2   && $-$          & $-$   & $<$1.5        & $< -$0.56
& $-$                           & W1a    & SNR?$ ^{e}$    \\
43.18+58.3  & 43.17+58.3$^{b}$  & u   & 5.5$\pm$0.1   && $-$          & $-$   & $<$1.5        & $< -$0.77
& $-$0.44$\pm$0.08$^{d}$        & $-$   & SNR$^{f}$     \\
43.21+61.3  & 43.21+61.3$^{a}$  & 0.8   & 4.3$\pm$0.1   && 43.25+61.4$^{a}$   & 1.3   & 4.0$\pm$0.3   & $-$0.04$\pm$0.05
& $-$                           & $-$   & HII?$^{e}$     \\
43.31+59.2  & 43.29+59.0$^{b}$  & u   & 8.0$\pm$0.2   && 43.30+59.2$^{b}$   & 0.4   & 2.0$\pm$0.2   & $-$0.83$\pm$0.06
& $-$0.65$\pm$0.07$^{d}$        & $-$   & SNR$^{f}$     \\
43.39+62.6  & 43.39+62.6$^{a}$  & 0.7   & 7.0$\pm$0.2   && 43.39+62.7$^{a}$   & 0.8   & 3.9$\pm$0.2   & $-$0.34$\pm$0.14 
&        $-$                    & Cd    &       \\
43.50+61.2  & 43.50+61.2$^{a}$  & 0.5   & 2.2$\pm$0.1   && $-$          & $-$   & $<$1.5        & $< -$0.23
&        $-$                    & $-$   &       \\
43.55+60.0  & 43.57+59.8$^{b}$  & 0.4   & 0.77$\pm$0.05 && $-$          & $-$   & $<$1.5        & $<$0.40
&        $-$                    & $-$   &       \\
43.65+57.7  & 43.65+57.7$^{a}$  & 0.7   & 2.3$\pm$0.1   && $-$          & $-$   & $<$1.5        & $< -$0.25
&        $-$                    & $-$   &       \\
44.01+59.6  & 44.02+59.5$^{b}$  & 0.4   & 19.5$\pm$0.2  && 44.01+59.5$^{b}$   & 0.4   & 4.5$\pm$0.4   & $-$0.87$\pm$0.05
& $-$0.38$\pm$0.06$^{d}$        & $-$   & SNR$^{f}$     \\
44.08+63.1  & 44.08+63.1$^{a}$  & 0.8   & 4.1$\pm$0.1   && $-$          & $-$   & $<$1.5        & $< -$0.60 
& $-$                           & $-$   &       \\
44.11+64.3  & 44.11+64.3$^{a}$  & 0.5   & 2.1$\pm$0.1   && $-$          & $-$   & $<$1.5        & $< -$0.20
& $-$                           & $-$   &  SNR?$^{e}$     \\
44.17+64.4  & $-$               & $-$   & $< 0.24$           && 44.17+64.4$^{a}$   & 0.8   & 4.1$\pm$0.2   & $> 1.74$
& $-$                           & $-$   &  HII$^{e}$     \\
44.29+59.3  & 44.28+59.3$^{b}$  & u   & 2.3$\pm$0.1   && $-$          & $-$   & $<$1.5        & $< -$0.25 
& $-$0.72$\pm$0.12$^{d}$        & $-$   & SNR$^{f}$     \\
44.36+57.8  & 44.36+57.8$^{a}$  & 0.4   & 1.5$\pm$0.1   && $-$          & $-$   & $<$1.5        & $<$0.0 
&       $-$                     & $-$   &       \\
44.43+62.5  & $-$               & $-$   & $< 0.24$      && 44.43+62.5$^{a}$   & 0.9   & 5.5$\pm$0.2   & $> 1.91$
&       $-$                     & E1b    &   HII$^{e}$    \\
44.50+65.3  & 44.50+65.3$^{a}$  & 0.5   & 1.4$\pm$0.1   && $-$          & $-$   & $<$1.5        & $<$0.0 
&       $-$                     & $-$   &       \\
44.52+58.1  & 44.52+58.0$^{b}$  & 0.4   & 2.1$\pm$0.05  && $-$          & $-$   & $<$1.5        & $< -$0.20
& $-$0.15$\pm$0.17$^{d}$        & $-$   & SNR$^{f}$     \\
44.84+61.8  & 44.84+61.8$^{a}$  & u   & 0.95$\pm$0.05 && $-$          & $-$   & $<$1.5        & $<$0.27
&       $-$                     & $-$   &       \\
44.91+61.1  & 44.89+61.2$^{b}$  & u   & 1.2$\pm$0.1   && $-$          & $-$   & $<$1.5        & $<$0.13
& $-$0.45$\pm$0.18$^{d}$        & $-$   & SNR$^{f}$     \\
44.93+63.9  & 44.92+63.7$^{b}$  & 0.7   & 3.6$\pm$0.1   && $-$          & $-$   & $<$1.5        & $< -$0.52
& $>$1.1$^{d}$                  & E2f    & HII$^{f}$     \\
45.17+61.2  & 45.19+61.1$^{b}$  & 0.4   & 6.7$\pm$0.1   && $-$          & $-$   & $<$1.5        & $< -$0.89
& $-$0.52$\pm$0.07$^{d}$        & $-$   & SNR$^{f}$     \\
45.38+60.3  & 45.38+60.3$^{a}$  & 0.4   & 1.1$\pm$0.05   && $-$          & $-$   & $<$1.5        & $<$0.19
& $-$                           & $-$   &       \\
45.44+67.3  & 45.44+67.3$^{b}$  & 0.4   & 1.47$\pm$0.06 && $-$          & $-$   & $<$1.5        & $<$0.0
& $-$                           & $-$   &       \\
45.63+66.9  & 45.63+66.9$^{a}$  & 0.8   & 5.8$\pm$0.2   && 45.61+67.1$^{a}$   & 0.9   & 5.5$\pm$0.4   & $-0.04$
& $-$                           & E2c    & HII$^{e}$     \\
45.70+62.9  & 45.66+62.8$^{b}$  & 0.4   & 0.78$\pm$0.05 && $-$          & $-$   & $<$1.5        & $<$0.39
& $-$                           & $-$   &       \\
45.79+65.2  & 45.74+65.5$^{b}$  & 0.5   & 5.9$\pm$0.2   && $-$          & $-$   & $<$1.5        & $< -$0.81
& $-$0.55$\pm$0.13$^{d}$        & E2b    & SNR$^{f}$     \\
45.91+63.8  & 45.88+63.7$^{b}$  & 0.4   & 1.28$\pm$0.05 && $-$          & $-$   & $<$1.5        & $<$0.01
& $-$0.38$\pm$0.11$^{d}$        & $-$   & SNR$^{f}$     \\
45.93+74.3  & 45.93+74.3$^{a}$  & 1.0   & 1.4$\pm$0.1   && $-$          & $-$   & $<$1.5        & $<$0.01
& $-$                           & $-$   &       \\
46.17+67.6  & 46.17+67.6$^{b}$  & 0.8   & 4.6$\pm$0.1   && 46.21+67.8$^{b}$   & 0.8   & 4.5$\pm$0.3   & -0.02$\pm$0.04
& 0.66$\pm$0.16$^{d}$           & E2a    & HII$^{f}$     \\
46.33+66.2  & 46.33+66.2$^{a}$  & u   & 0.93$\pm$0.05 && $-$          & $-$   & $<$1.5        & $<$0.28
& $-$                           & $-$   &       \\
46.52+63.8  & 46.53+63.9$^{b}$  & 0.4   & 2.1$\pm$0.1   && $-$          & $-$   & $<$1.5        & $< -$0.20
& $-$0.35$\pm$0.11$^{d}$        & $-$   & SNR$^{f}$     \\
46.56+73.8  & 46.57+73.7$^{b}$  & u   & 0.76$\pm$0.05 && $-$          & $-$   & $<$1.5        & $<$0.40
& $-$0.60$\pm$0.05$^{c}$        & $-$   & SNR$^{f}$     \\
46.74+69.7  & 46.74+69.7$^{a}$  & 1.4   & 4.3$\pm$0.2   && $-$          & $-$   & $<$1.5        & $< -$0.63
&        $-$                    & $-$   & SNR?$^{e}$     \\
46.75+67.0  & 46.69+66.9$^{b}$  & 0.4   & 1.8$\pm$0.1   && $-$          & $-$   & $<$1.5        & $<$0.11
& $-$0.80$\pm$0.05$^{c}$        & $-$   & SNR$^{f}$     \\
47.11+66.3  & 47.11+66.3$^{a}$  & 0.5   & 1.12$\pm$0.05 && $-$          & $-$   & $<$1.5        & $<$0.17
&        $-$                    & $-$   &       \\
47.37+68.0  & 47.37+67.8$^{b}$  & 0.4   & 0.7$\pm$0.05  && $-$          & $-$   & $<$1.5        & $<$0.45
& $-$0.60$\pm$0.07$^{c}$        & $-$   & SNR$^{f}$     \\

\enddata
\tablenotetext{1}{Relative to (B1950) 09$^{h}$51$^{m} +$ 69$^{\circ}$ 54$'$.}
\tablenotetext{2}{Deconvolved angular size from the 0.6$''$ HPBW. With this resolution the
sources which appear unresolved are indicated by "u".}
\tablenotetext{a}{Sources newly identified in this paper.}
\tablenotetext{b}{From \citet{Kr85,Ba87,Hu94,Mu94,Wi97,Mc02}.}
\tablenotetext{c}{Spectral index ($S\propto\nu^{\alpha}$) between 1.4, 5 and 8.4~GHz \citep{Wi97}.}
\tablenotetext{d}{Spectral index ($S\propto\nu^{\alpha}$) between 5 and 15~GHz \citep{Mc02}.}
\tablenotetext{e}{Source type obtained in this paper based on the spectral index in column 8.}
\tablenotetext{f}{Source type given by \citet{Mc02}.}
\end{deluxetable}

\begin{deluxetable}{cccccccccccc}
\tabletypesize{\scriptsize}
\tablecolumns{10}
\tablewidth{0pc}
\tablecaption{
L{\footnotesize INE PARAMETERS AT LOW ANGULAR RESOLUTION USED IN THE MODELS.} }
\tablehead{
  \colhead{}
& \multicolumn{4}{c}{H92$\alpha$}
& \colhead{} &\multicolumn{4}{c}{H53$\alpha$}
\\ \cline{2-4}  \cline{7-9} \\
  \colhead{Region}

& \colhead{Peak Flux}
& \colhead{V$_{HEL}$}
& \colhead{$\Delta$V}
& \colhead{S$_{C8}$}
&
& \colhead{Peak Flux}
& \colhead{V$_{HEL}$}
& \colhead{$\Delta$V}
& \colhead{S$_{C43}$}
& \colhead{ff$^{b}$}
& \colhead{S$_{C5}$}
\\
  \colhead{}
& \colhead{(mJy)}
& \colhead{(km/s)}
& \colhead{(km/s)}
& \colhead{(mJy)}
&
& \colhead{(mJy)}
& \colhead{(km/s)}
& \colhead{(km/s)}
& \colhead{(mJy)}
& \colhead{(10$^{-3}$)}
& \colhead{(mJy)}
}
\startdata
E2      &   2.2$\pm$0.1  &   315$\pm$2   &   106$\pm$5   & 106$\pm$10 &
        &   7.5$\pm$1.5  &   305$\pm$5   &   55$\pm$12   & 46$\pm$17   & 1.3  & 152 \\

E1      &   1.0$\pm$0.1  &   269$\pm$3   &   130$\pm$8   & 72$\pm$10   &
        &   6.7$\pm$1.5  &   264$\pm$5   &   50$\pm$13   & 28$\pm$15   & $--$ & 93 \\

W1      &   6.4$\pm$0.1  &   119$\pm$1   &   116$\pm$3   & 292$\pm$15  &
        &   9.8$\pm$2.0  &   113$\pm$12  &   120$\pm$29  & 157$\pm$20  & 3.2  & 403 \\

W2      &   4.2$\pm$0.1  &   100$\pm$2   &   112$\pm$4   & 169$\pm$15  &
        &   10.2$\pm$2.0 &   66$\pm$9    &   100$\pm$22  & 101$\pm$19  & 6.8  & 230 \\

\enddata
\tablenotetext{a}{Note that the integrated line flux density for each region is lower than
that given in Table~4 (column 2) since the regions for which we have integrated here are defined
by the lowest contour in the H53$\alpha$ integrated line emission image.}
\tablenotetext{b}{Area filling factor computed using Table~4 from \citet{Mc02}.}
\end{deluxetable}

\hoffset -0.5in
\begin{deluxetable}{cccccccccc}
\tabletypesize{\scriptsize}
\tablecolumns{10}
\tablewidth{0pc}
\tablecaption{
P{\footnotesize ARAMETERS FOR REGIONS OBSERVED IN H92$\alpha$ RRL}  }
\tablehead{
  \colhead{}
& \multicolumn{2}{c}{Low angular resolution}
& \colhead{} &\multicolumn{6}{c}{Intermediate resolution}
\\ \cline{2-3}  \cline{5-10} \\

  \colhead{Region}
& \colhead{$\int$S$_{L}$dV}
& \colhead{S$_{L}$/S$_{C}^{a}$}
& \colhead{Features}
& \colhead{Coordinates$^{b}$}
& \colhead{$\int$S$_{L}$dV}
& \colhead{Size$^{c}$}
& \colhead{Peak Flux}
& \colhead{V$_{HEL}$}
& \colhead{$\Delta$V$^{d}_{FWHM}$}

\\

  \colhead{}
& \colhead{(mJy~km~$s^{-1}$)}
& \colhead{($\times10^{-2}$)}
&
&
& \colhead{(mJy~km~$s^{-1}$)}
& \colhead{$('')$}
& \colhead{(mJy)}
& \colhead{(km/s)}
& \colhead{(km/s)} \\

\colhead{(1)}   & \colhead{(2)} & \colhead{(3)} & \colhead{(4)}         & \colhead{(5)} & \colhead{(6)} & \colhead{(7)} &
 \colhead{(8)}  & \colhead{(9)} & \colhead{(10)}        \\

}

\startdata
 E2     &   604      &       1.7$\pm$0.1     &       E2a      &       46.20+67.7     &       81
        &   1.3      &       0.57$\pm$0.03   &       316$\pm$16       & 121$\pm$37      \\

        &            &                       &       E2b      &       45.79+65.3     &       65
        &   1.4      &       0.82$\pm$0.05   &       315$\pm$5        & 48$\pm$11    \\

        &            &                       &       E2c      &       45.54+66.3     &       165
        &   1.6      &       1.54$\pm$0.14   &       320$\pm$8        & 84$\pm$19       \\

        &            &                       &       E2d      &       45.43+65.6     &       96
        &   0.8      &       0.82$\pm$0.05   &       311$\pm$5        & 95$\pm$11    \\

        &            &                       &       E2e      &       45.32+64.2     &       69
        &   0.9      &       0.61$\pm$0.05   &       314$\pm$5        & 90$\pm$11    \\

        &            &                       &       E2f      &       45.05+63.5     &       32
        &   0.4      &       0.26$\pm$0.06   &       302$\pm$7        & 102$\pm$16      \\

        &            &                       & E2a+E2b+...+E2f &               &       508                \\

 \cline{1-10} \\

E1      &   300         &       1.2$\pm$0.1       &       E1a      &       44.39+61.8     &       98
        &   1.4         &       0.87$\pm$0.07     &       264$\pm$8       &   90$\pm$18       \\

        &               &                         &       E1b      &       44.10+62.7     &       56
        &   1.6         &       0.56$\pm$0.05     &       321$\pm$10      &   76$\pm$24       \\

        &               &                         &       E1c      &       44.02+60.3     &       39
        &   1.4         &       0.59$\pm$0.05     &       231$\pm$10      &   24$\pm$24       \\

        &               &                       &   E1a+E1b+E1c     &       &       193          \\

\cline{1-10} \\

C       &   320    &        1.0$\pm$0.1      &        Ca     &       43.84+62.4     &       70
        & 0.9      &       0.63$\pm$0.05     &       300$\pm$5       &       88$\pm$11       \\

        &          &                         &        Cb     &       43.77+63.2     &       137
        & 1.2      &       0.61$\pm$0.05     &       286$\pm$5       &       205$\pm$11       \\

        &          &                         &        Cc     &       43.79+63.9     &       16
        & 0.6      &       0.30$\pm$0.05     &       470$\pm$5       &       52$\pm$11       \\

        &          &                         &        Cd     &       43.39+62.3     &       75
        & 1.3      &       1.00$\pm$0.05     &       239$\pm$4       &       44$\pm$9        \\

        &               &                       &   Ca+Cb+Cc+Cd     &       &       298          \\

\cline{1-10} \\

W1       &  1067     &        1.6$\pm$0.1    &   W1a      &       42.95+59.2     &       85
        & 1.9       &       0.72$\pm$0.10    &   165$\pm$8       &  96$\pm$4        \\

        &           &                        &   W1b      &       42.62+60.3     &       196
        & 2.1       &       1.85$\pm$0.10    &   148$\pm$3       &  83$\pm$8        \\

        &           &                        &   W1c      &       42.61+58.0     &       291
        & 1.7       &       2.88$\pm$0.07    &   117$\pm$2       &  77$\pm$3        \\

        &           &                        &   W1d      &       42.32+61.5     &       57
        & 1.1       &       0.56$\pm$0.10     &  191$\pm$8       &  78$\pm$4        \\

        &           &                        &   W1e      &       42.21+59.9     &       147
        & 1.3       &       1.05$\pm$0.07    &   125$\pm$2       &  120$\pm$3        \\

        &           &                        &   W1f      &       42.19+58.8     &       177
        & 1.6       &       2.01$\pm$.10     &   100$\pm$3        & 61$\pm$8        \\

        &           &                        &  W1a+W1b+...+W1f    &       & 953  &              \\

\cline{1-10} \\

W2      &   878         &      2.0$\pm$0.1      &       W2a     &       41.75+59.0     & 113
        & 1.4      &       0.76$\pm$0.07     &       114$\pm$12        &  128$\pm$32      \\

        &               &                       &       W2b     &       41.62+57.8     & 141
        & u        &       1.43$\pm$0.07     &       79$\pm$12        &   69$\pm$32      \\

        &               &                       &       W2c     &       41.51+58.6     & 81
        & 1.2      &       0.68$\pm$0.07     &       113$\pm$3        &   96$\pm$6      \\

        &               &                       &       W2d     &       41.47+56.6     & 69
        & 1.6      &       0.80$\pm$0.08     &       76$\pm$5        &    57$\pm$12        \\

        &               &                       &       W2e     &       41.27+59.2     & 65
        & 1.1      &         0.69$\pm$0.11     &     135$\pm$5       &    68$\pm$10       \\

        &               &                       &       W2f     &       41.12+56.4     & 113
        & 1.8      &       1.19$\pm$0.08     &       86$\pm$5        &    69$\pm$12        \\

        &               &                       &       W2g     &       41.07+61.0     & 36
        & 1.3      &       0.23$\pm$0.07     &       84$\pm$3        &    133$\pm$6      \\

        &               &                       &       W2h     &       40.97+58.5     & 112
        & 1.4      &         1.14$\pm$0.09     &     129$\pm$4       &    73$\pm$9       \\

        &          &                           & W2a+W2b+...+W2h  &      &       730                 \\

\cline{1-10} \\

\enddata
\tablenotetext{a}{The S$_{C}$ and S$_{L}$ values are listed in Table~3.}
\tablenotetext{b}{Relative to (B1950) 09$^{h}$51$^{m} +$ 69$^{\circ}$ 54$'$.}
\tablenotetext{c}{Deconvolved angular size from $0\rlap.{''}9$ synthesized beam, u is given when the source is unresolved.}
\tablenotetext{d}{Deconvolved line width from 56~km~$^{-1}$, the spectral resolution achieved in the H92$\alpha$ line.}
\end{deluxetable}

\clearpage

\begin{deluxetable}{cccccccccc}\label{tab5}
\tablewidth{0pt}
\tabletypesize{\scriptsize}
\tablecaption{M{\footnotesize ODELS}
B{\footnotesize ASED ON THE} H92$\alpha$ {\footnotesize AND} H53$\alpha$ L{\footnotesize INES}}
\tablehead{
\colhead{Parameter}     & \multicolumn{2}{c}{Region E2}  & 
\multicolumn{1}{c}{Region E1} & \multicolumn{2}{c}{Region W1} & \multicolumn{2}{c}{Region W2} \\
                        & \colhead{SD\tablenotemark{d}} & \colhead{TD\tablenotemark{d}}  & 
			\colhead{SD\tablenotemark{d}} & \colhead{SD\tablenotemark{d}} & \colhead{TD\tablenotemark{d}} 
			& \colhead{SD\tablenotemark{d}} & \colhead{TD\tablenotemark{d}} }
\startdata
 T$_{e} \times 10 ^{3}$ (K) \dotfill &
  7$\pm$2     &  $\big\{$$7.5\pm2.5 \atop 6\pm1$ &
  7.5$\pm$2.5 & 
  6$\pm$0.5   &  $\big\{$$7.5\pm2.5 \atop 5\pm0.5$ &
  7.5$\pm$2.5 &  $\big\{$$7.5\pm2.5 \atop 6\pm1$ 
\vspace{0.15cm}
\\
 n$_{e} \times10^{3}$(cm$^{-3}$) \dotfill &
  9$\pm$8     &  $\big\{$$40\pm20 \atop 5\pm4$ &
  26$\pm$25   &
  4$\pm$3     &  $\big\{$$30\pm10 \atop 4.0\pm2.5$ &
  15$\pm$13   &  $\big\{$$30\pm10 \atop 5.5\pm4.0$ 
\vspace{0.15cm}
\\
 Size (pc)\dotfill &
  0.81$\pm$0.59 &   $\big\{$$0.14\pm0.05 \atop 0.85\pm54$ &
  0.75$\pm$0.65 &
  0.72$\pm$0.34 &   $\big\{$$0.15\pm0.04 \atop 0.73\pm33$ &
  0.52$\pm$0.36 &   $\big\{$$0.15\pm0.04 \atop 0.69\pm0.38$ 
\vspace{0.15cm}
\\
 N$_{HII} \times 10 ^{3}$ \dotfill &
  2.5$\pm$1.0  &   $\big\{$$0.56\pm0.35 \atop 2.4\pm1.1$ &
  1.7$\pm$0.88 & 
  5.8$\pm$1.1  &   $\big\{$$1.6\pm0.7 \atop 4.6\pm0.3$ &
  6.8$\pm$3.8  &   $\big\{$$2.4\pm1.0 \atop 4.6\pm2.6$ 
\vspace{0.15cm}
\\
 f\tablenotemark{a} \dotfill &
 0.29$\pm$0.28 &   $\big\{$$0.0013 \atop 0.29\pm0.27$ &
 0.27$\pm$0.26 &  
 0.31$\pm$0.24 &   $\big\{$$0.0032 \atop 0.30\pm0.23$ &
 0.37$\pm$0.35 &   $\big\{$$0.0068 \atop 0.40\pm0.37$ 
\vspace{0.15cm}
\\
Log N$_{Ly}$~(s$^{-1}$) \dotfill &
52.36$\pm$0.19 &   $\big\{$$51.64\pm0.32 \atop 52.33\pm0.21$ &
52.16$\pm$0.25 & 
52.75$\pm$0.09 &   $\big\{$$52.15\pm0.2 \atop 52.67\pm0.03$ &
52.75$\pm$0.27 &   $\big\{$$52.34\pm0.2 \atop 52.58\pm0.27$ 
\vspace{0.15cm}
\\
 SFR\tablenotemark{b}~(M$_{\odot}$~year$^{-1}$) \dotfill &
0.37$\pm$0.13 &   $\big\{$$0.10\pm0.07 \atop 0.45\pm0.19$ &
0.21$\pm$0.11 & 
1.64$\pm$0.13 &   $\big\{$$0.29\pm0.13 \atop 0.86\pm0.05$ &
0.63$\pm$0.51 &   $\big\{$$0.45\pm0.19 \atop 0.86\pm0.48$ 
\vspace{0.15cm}
\\
 $\tau_{c}$ (8.3 GHz)\dotfill &
 0.27$\pm$0.26  &   $\big\{$$0.032\pm0.023 \atop 0.14\pm0.13$ &
 0.51$\pm$0.50  &  
 0.10$\pm$0.08  &   $\big\{$$0.015\pm0.007 \atop 0.09\pm0.07$ &
 0.35$\pm$0.32  &   $\big\{$$0.015\pm0.007 \atop 0.14\pm0.12$ 
\vspace{0.15cm}
\\
 $\tau_{L}$ (8.3 GHz)\dotfill &
$-0.13 \pm 0.10$  &   $\big\{$$-0.050\pm0.048 \atop -0.095\pm0.068$ &
$-0.16 \pm 0.13$  &
$-0.09 \pm 0.05$   &   $\big\{$$-0.032\pm0.027 \atop -0.085\pm0.042$ &
$-0.14\pm 0.11$   &   $\big\{$$-0.051\pm0.047 \atop -0.095\pm0.068$ 
\vspace{0.15cm}
\\
 b$_{n}$ (H92$\alpha$)\dotfill &
0.97$\pm$0.02 &   $\big\{$$0.92\pm0.03 \atop 0.96\pm0.02$ &
0.97$\pm$0.03 &  
0.97$\pm$0.02 &   $\big\{$$0.91\pm0.025 \atop 0.97\pm0.014$ &
0.98$\pm$0.02 &   $\big\{$$0.91\pm0.025 \atop 0.97\pm0.02$ 
\vspace{0.15cm}
\\
 $ \beta_{n}$ (H92$\alpha$)\dotfill &
$-20 \pm$15 &   $\big\{$$-23\pm8.1 \atop -22\pm13$ &
$-20 \pm$16 &  
$-19 \pm$9  &   $\big\{$$-26\pm5 \atop -20\pm8$ &
$-18 \pm$12 &   $\big\{$$-26\pm5 \atop -20\pm11$ 
\vspace{0.15cm} 
\\
 $\alpha$\tablenotemark{c}\dotfill &
0.95$\pm$0.20  &   $ 1.0\pm0.24$ &
1.00$\pm$0.30  &
0.50$\pm$0.03  &   $ 0.52\pm0.026 $ &
0.91$\pm$0.5   &   $ 0.62\pm0.19$ 
\vspace{0.15cm}
\\
 M$_{HII} \times 10 ^{4}$ (M$_{\odot}$)\dotfill &
2.5$\pm$2.6  &   $\big\{$$0.041\pm0.008 \atop 3.0\pm2.6$ &
1.5$\pm$1.4  &  
6.6$\pm$4.5  &   $\big\{$$0.16\pm0.02 \atop 6.1\pm4.0$ &
5.6$\pm$5.1  &   $\big\{$$0.25\pm0.03 \atop 6.1\pm5.4$ 
\vspace{0.15cm}
\\
\enddata
\tablenotetext{a}{Area covering factor.}
\tablenotetext{b}{Star formation rate using the formulae given by \citet{An00}.}
\tablenotetext{c}{Spectral index between 8.3 and 43~GHz.}
\tablenotetext{d}{Results from single-density models are listed in columns SD and results from two-density models are listed in columns labeled TD.}
\end{deluxetable}

\clearpage

\begin{deluxetable}{cccccccccc}\label{tab6}
\tablewidth{0pt}
\tabletypesize{\scriptsize}
\tablecaption{M{\footnotesize ODELS FOR A NUCLEAR RING.}}
\tablehead{
\colhead{Ring center}$^{b}$  & \colhead{PA}$^{b}$    & \colhead{Maj}$^{b}$ & \colhead{Min}$^{b}$ & \colhead{Vsys}$^{b}$ & \colhead{Vrot}$^{b}$ & 
\colhead{Vexp}$^{b}$ & \colhead{$\sigma_{fit}$} & Line \\
\colhead{    }         & \colhead{(deg.) }&\multicolumn{2}{c}{(arcsec)}  & \multicolumn{4}{c}{(km s$^{-1}$)} & }
\startdata
{\bf 9$^{h}$51$^m$43$\rlap.{^{s}}$40, 69$^{\circ}$55$'$0$\rlap.{''}$1}$^{b}$& {\bf 70} & {\bf 11.4} & {\bf 3.3} & 200 & 112 & -16   &      & [Ne~II]$^a$\\
{\bf 9$^{h}$51$^m$43$\rlap.{^{s}}$40, 69$^{\circ}$55$'$0$\rlap.{''}$1}$^{b}$& {\bf 70} & {\bf 11.4} & {\bf 3.3} & 193 &  94 & -13.5 & 7.47 &    H92$\alpha$ \\
{\bf 9$^{h}$51$^m$43$\rlap.{^{s}}$40, 69$^{\circ}$55$'$0$\rlap.{''}$4}$^{b}$& {\bf 70} & {\bf 11.4} & {\bf 3.3} & 196 &  96 & -15   & 7.27 &    H92$\alpha$ \\
     9$^h$51$^m$43$\rlap.{^{s}}$19, 69$^{\circ}$55$'$0$\rlap.{''}$6 & {\bf 70} & {\bf 11.4} & {\bf 3.3} & 191 & 102 & -13  & 7.35 &    H92$\alpha$ \\
     9$^h$51$^m$43$\rlap.{^{s}}$28, 69$^{\circ}$55$'$0$\rlap.{''}$6 &      66  & {\bf 11.4} &      5.6  & 195 & 103 & -14  & 4.82 &    H92$\alpha$ \\
\enddata
\tablenotetext{a}{Results from \citet{Ach95}.}
\tablenotetext{b}{The fixed parameters are given in bold face and free parameters estimated from the fit in normal face.}
\end{deluxetable}

\clearpage

\voffset -0.25in

\clearpage

\clearpage

\clearpage

\clearpage

\end{document}